\documentclass[11pt]{article}
\usepackage{aaspp4,lscape}

\def\gtsima{$\, \buildrel > \over \sim \,$}
\def\ltsima{$\, \buildrel < \over \sim \,$}
\def\simgt{\lower.5ex\hbox{\gtsima}}
\def\simlt{\lower.5ex\hbox{\ltsima}}

\begin{document}

\title{ Calibration of the MACHO Photometry Database }

\author{
      C.~Alcock\altaffilmark{1,16}, 
      R.A.~Allsman\altaffilmark{2},
      D.R.~Alves\altaffilmark{1,3,4},
      T.S.~Axelrod\altaffilmark{5},
      A.C.~Becker\altaffilmark{6},\\
      D.P.~Bennett\altaffilmark{4,7,16},
      K.H.~Cook\altaffilmark{1,4,16},
      A.J.~Drake\altaffilmark{5},
      K.C.~Freeman\altaffilmark{5},
      K.~Griest\altaffilmark{8,16},
      M.J.~Lehner\altaffilmark{9},\\
      S.L.~Marshall\altaffilmark{1,4},
      D.~Minniti\altaffilmark{1,10}, 
      B.A.~Peterson\altaffilmark{5},
      M.R.~Pratt\altaffilmark{11}, 
      C.A.~Nelson\altaffilmark{12},
      P.J.~Quinn\altaffilmark{13},\\
      C.W.~Stubbs\altaffilmark{6,16}, 
      W.~Sutherland\altaffilmark{11,16}, 
      A.B.~Tomaney\altaffilmark{6},
      T.~Vandehei\altaffilmark{8},
      D.L.~Welch\altaffilmark{15}
}

\begin{center}
{\bf (The MACHO Collaboration)}
\end{center}

\vspace{-10mm}
{\footnotesize
\altaffiltext{1}{Lawrence Livermore National Laboratory, Livermore, CA 94550}
\altaffiltext{2}{Supercomputing Facility, Australian National University,
        Canberra, ACT 0200, Australia }
\altaffiltext{3}{Department of Physics, University of California, Davis, CA 95616;
now affiliated with the Space Telescope Science Institute, Baltimore, MD 21218}
\altaffiltext{4}{Visiting Astronomer, Cerro Tololo
Inter-American Observatory, which is operated by the Association of
Universities for Research in Astronomy, Inc., under cooperative agreement with
the National Science Foundation.}
\altaffiltext{5}{Mt.~Stromlo and Siding Spring Observatories, ANU,
        Weston Creek, ACT 2611, Australia}
\altaffiltext{6}{Departments of Astronomy \& Physics,
        University of Washington, Seattle, WA 98195 }
\altaffiltext{7}{Physics Department, University of Notre Dame, Notre 
        Dame, IN 46556 }
\altaffiltext{8}{Department of Physics, University of California,
        San Diego, La Jolla, CA 92093 }
\altaffiltext{9}{Department of Physics, University of Sheffield,
        Sheffield S3 7RH, UK }
\altaffiltext{10}{Departmento de Astronomia, 
P. Universidad Catolica, Casilla 104, Santiago 22, Chile}
\altaffiltext{11}{Center for Space Research, MIT, Cambridge MA 02139 }
\altaffiltext{12}{Department of Physics, University of California,
        Berkeley, CA 94720 }
\altaffiltext{13}{European Southern Observatory, Karl-Schwarzchild Str. 2,
        D-85748, Garching, Germany }
\altaffiltext{14}{Department of Physics, University of Oxford,
        Oxford OX1 3RH, U.K. }
\altaffiltext{15}{Departments of Physics \& Astronomy, 
   McMaster University, Hamilton, Ontario, Canada L8S 4M1. }
\altaffiltext{16}{Center for Particle Astrophysics,
        University of California, Berkeley, CA 94720}
}

\clearpage

\begin{abstract}

The MACHO Project is a microlensing survey that monitors
the brightnesses of $\sim$60 million stars 
in the Large Magellanic Cloud (LMC), Small
Magellanic Cloud, and Galactic bulge.  Our database presently
contains about 80 billion photometric measurements, a
significant fraction of all astronomical photometry.
We describe the calibration of MACHO
two-color photometry and transformation to the standard 
Kron-Cousins $V$ and $R$ system.  Calibrated MACHO photometry
may be properly compared with all other observations
on the Kron-Cousins standard system, enhancing the astrophysical
value of these data.
For $\sim$9 million stars in the LMC bar,
independent photometric measurements of $\sim$20,000
stars with $V \simlt 18$ mag in field-overlap
regions demonstrate an internal precision $\sigma_{V} = 0.021$,
$\sigma_{R} = 0.019$, $\sigma_{V-R} = 0.028$ mag.
The accuracy of the zero-point in this calibration is estimated
to be $\pm 0.035$ mag for stars with colors 
in the range $-0.1 < (V-R) < 1.2$ mag.
A comparison of calibrated MACHO photometry with published
photometric sequences and new {\it Hubble Space Telescope}
observations shows agreement.
The current calibration zero-point uncertainty for the remainder
of the MACHO photometry database
is estimated to be $\pm 0.10$ mag in $V$ or $R$ and $\pm 0.04$ mag in $(V-R)$.
We describe the
first application of calibrated MACHO photometry data: the construction
of a color-magnitude diagram used to
calculate our experimental sensitivity to detect microlensing
in the LMC.

\end{abstract}


\section{ Introduction }

The MACHO Project is a microlensing survey experiment (Alcock et al.~1997) 
that monitors the brightness variations of $\sim$60 million stars in the
Large Magellanic Cloud (LMC), Small Magellanic Cloud (SMC) and Galactic
bulge.  Microlensing is the rare transient magnification of a background source 
star due to the gravitational effect of a massive
compact object crossing the line of sight.  
Paczy\'{n}ski (1986) first noted that if the dynamically-inferred
Galactic dark halo 
was composed of massive compact objects, the probability of
microlensing would be $\tau \sim 5 \times 10^{-7}$
toward the LMC,
within reach of dedicated observational surveys.
The MACHO Project survey observations are
made with a mosaic of charge-coupled devices imaging simultaneously
in non-standard blue and red passbands.  The special purpose
instrument is permanently mounted on the 50-inch Great Melbourne
Telescope
in Australia (Hart et al.~1996).  
The total sky area monitored is approximately 40, 3,
and 45 square degrees in the LMC, SMC, and Galactic bulge, respectively.  
Each star is represented in the MACHO database by a time-series of two-color
photometric measurements.  In some cases, stars are counted in the database 
two or three times because the survey fields overlap on the sky.

In this paper, we describe the calibration of MACHO photometry data.
Calibration actually encompasses several levels of detail regarding 
the systematic transformation of the MACHO data to a meaningful absolute
system.  The first level of calibration
is the creation of an instrumental system. 
The second level of calibration is the transformation of the instrumental photometry
to the Kron-Cousins $V$ and $R$ standard system\footnote{For an excellent discussion
of different optical broad-band photometric systems, the
reader is referred to Bessell (1979, 1986, 1987, 1990, 1995).}.
This level of calibration allows
for proper comparison of MACHO data with all other data on this system.
In practice, these two levels of calibration are not implemented separately.
The third level of calibration discussed in this paper is the
calibration of lightcurves, i.e.~analysis-stage corrections that may
be applied to time-series photometric data for individual stars.   
These corrections will eliminate some systematic sources of scatter
in the photometry.
Our first effort is to calibrate the ``top-22'' LMC
fields analysed in Alcock et al.~(1997).

The MACHO Project 
microlensing analyses involve the photometric calibration in a number 
of ways.  First,
regions of the LMC color-magnitude diagram
(CMD) excluded in the search for microlensing because of the high
background of intrinsic variable stars are more accurately defined
with the calibrated data.  It is also possible to
make a more precise comparison of the distribution of microlensing
source stars in the CMD with that expected for true microlensing.  
Finally, calibration plays an important role in the
analysis of LMC microlensing through the calculation of our
experimental sensitivity to detect microlensing, 
which we will call the ``efficiency.'' 

The microlensing efficiency calculation
requires a critical assumption regarding the true distribution
of stars in the LMC for the following reason.  
Individual stars in the ground-based MACHO
image data are almost always confused, i.e. they are
composites of two or more unresolved companions.  
However, only one true LMC star will be lensed. 
The partial magnification of flux from an unresolved,
composite star is an effect  
known as blending in microlensing.
Observational data and further
discussion of blending can be found in Alcock et al.~(1997).
It is useful to distinguish an apparent star in the MACHO data
which is actually a blend of several real ``stars'' by referring to 
it as an ``object.''  
The efficiency calculation (an exhaustive series of artifical star tests and
Monte-Carlo experiments) returns our experimental sensitivity 
to detect microlensing of stars,
not objects.

In order to quantify the ratio of stars to objects, we have obtained
{\it Hubble Space Telescope} (HST) imaging data in their filter equivalents
of $V$ and $R$ for three fields in the LMC top-22.  The high spatial resolution
of the HST allows us to probe to fainter magnitudes than possible with our
ground-based data, particularly in the crowded bar region.  
The HST and MACHO data are properly comparable
after their respective calibration to the Kron-Cousins
standard $V$ and $R$ magnitude system.
We construct the ``efficiency CMD'',
which is a properly scaled splicing of MACHO and HST data together into
a CMD that also contains accurate information on the numbers of stars.
The scaling factor is essentially the sought-after
star to object ratio.  The efficiency CMD described here has been
used to seed millions of artificial stars into our raw image data,
and is the first application of calibrated MACHO data.
Complete discussions of the new LMC microlensing analysis, the
efficiency calculation, and the HST data reduction are
beyond the scope of this paper; each will be the subject
of a forthcoming paper.

In addition to microlensing, 
the MACHO photometry database is a valuable resource for studying
the stellar populations and star formation
history of the LMC, making new tests of stellar evolution theory, and 
for studying variable stars.  
Calibration
enhances the value of the MACHO database
for these so-called science ``by-products.'' 
For this reason, {\it we give special attention to
details of the calibrations that may be relevant
to consumers of released MACHO data}.

Our paper is organized as follows.  
In \S2, we preface the calibration discussion with
some details of the MACHO image and photometry data. 
In \S3, we review the calibration of the LMC top-22 fields.
In \S4, we compare the calibrated MACHO data with a sample of
published photometric sequences in the LMC and with the
new HST observations.
In \S5, we describe the status of calibration for the 
remaining fields in the LMC, SMC, and Galactic bulge.
In \S6, 
we review the calibration of MACHO instrumental 
lightcurves.  
In \S7, we examine the HST and MACHO data in greater
detail and construct the efficiency CMD.
Finally, \S8 is the summary of our results.

\section{ MACHO Data }

\subsection{ Images }

The MACHO experiment 
has dedicated use of the 1.27-m (50-inch) Great Melbourne Telescope
(Robinson \& Grubb 1869),
now located at the Mount Stromlo Observatory in Australia.
A system of corrective optics has been installed
at the prime focus, giving a focal reduction to $f/3.9$ and
a $1^{\circ}$ field of view.  
A dichroic beam-splitter 
enables simultaneous blue and red imaging
(Hart et al.~1996).  
Observations may be made from either side
of the telescope pier (a German equatorial mount),
i.e. either an East or West of pier observation.
The median stellar image FWHM is 2~arcsec.
The typical sky for an LMC bar field is estimated to
be $R$ $\sim$ 19.5 mag per square arcsec\footnote{This estimate
likely includes some contribution from an unresolved stellar background
in the LMC bar.  It is calculated from $\sim$200
images of one field spanning the full range of conditions encountered
over four years of observations.}.

The MACHO
filters were specially designed to provide an adequate color
index and wide bandpasses.  
The blue filter runs from $\sim$4500$-$6300 $\AA$ 
and the red filter runs from $\sim$6300$-$7600
$\AA$.
At both the red and blue foci, a mosaic of four
2048$\times$2048 Loral
charge coupled devices (CCDs) are mounted.  
In Figure 1, we
show the approximate response of the dichroic, filters,
and CCDs.  The wide-field optics corrector has not been included.
Uncertainty is estimated to be $\sim$20\% in these response functions.
The normalized standard $BVRI$ passbands from
Bessell (1990) are also shown.

The Loral CCD pixel size is $15\, \mu m$ which corresponds
to $0.''635$ on the sky, giving a sky coverage
of $0.52$ square degrees per MACHO field.  
Each CCD has two read-out amplifiers,
and the images are read-out through a 16-channel system and
written into dual-ported memory in the data acquisition computer.
The readout time is 67 seconds per image, and the noise is
$\sim 10$ electrons rms, with a gain of $\sim 2 \, e^-$/ADU.
Further details of the MACHO camera system are provided
in Stubbs et al.~(1993) and Marshall (1994).

In Figure 2, we present a schematic 
drawing of the red and blue MACHO focal planes 
from both the East and West sides of the pier.  
We label each CCD (0-3 on the red 
side, and 4-7 on the blue side), and each amplifier (.0 or .1).
The CCD-amplifier designated 0.0
is inoperative;
it is marked with an ``X.''
For the purposes of photometry, 
each red and blue MACHO image 
is divided into 64 ``chunks'' (128 total),
each approximately $512 \times 512$ pixels in size.
Chunks are defined as regions of a certain CCD
and amplifier in the MACHO focal plane\footnote{Template
photometry for certain chunks may be derived from a different 
CCD-amplifier image than its 
otherwise ``defined'' location in the focal plane.}.
Every red chunk uniquely corresponds to a blue chunk.
In the bottom two panels of Figure 2, we present 
chunk maps in the West of pier orientation.

\subsection{ SoDOPHOT }

The photometry for the MACHO experiment is handled by a special purpose
code called SoDOPHOT, which stands for ``Son of DOPHOT.''  
The reader is referred to 
Schecter, Mateo, and Saha (1993) for further
details of DOPHOT.  Briefly, it is a model-based fitting code that
searches for objects in two dimensional digital array images of the sky.
Stars, galaxies, cosmic rays, and so forth are each assigned a specific model
defined in terms of analytic functions.  Objects are identified and photometered
with precise signal-to-noise criteria based on the model fits.
DOPHOT may tend to report brighter
magnitudes for faint stars in crowded regions than Daophot
(Stetson 1987), a similar type of photometry code that is 
widely employed by astronomers.  This systematic effect 
is attributed to the sky fitting procedure (Schecter et al. 1993),
and is likely preserved in SoDOPHOT.

SoDOPHOT is basically DOPHOT optimized 
to the MACHO image data and modified for
extremely fast CPU reduction times.  
Most of the improvement in speed can be attributed to the fact
that we observe the same fields repeatedly. 
We designate a high quality image of a given field as a ``template" image
and use the reduction of our template image to help with the reduction of the
other images.  The template starlists
are generated by an iterative and automated
DOPHOT reduction which employs
both the red and blue MACHO template images.

For routine reductions, SoDOPHOT makes a 
list of stellar positions and brightnesses of the brightest stars in the
new image and finds a crude transformation to the positions and relative
brightness of stars from the template. 
This transformation is then used to find the approximate location of
a set of about 40 bright and relatively isolated fiducial stars per chunk
which have been preselected from the list of template stars. 
SoDOPHOT reductions are done on individual chunks, i.e.
$\sim$1/16 of a Loral CCD and $\sim$1/64 of the imaged area in the focal plane.
The fiducial stars are
then subject to a 7-parameter fit to find their precise position, brightness,
and point spread function (PSF) shape. The PSF model is a 3-parameter 
``pseudo-elliptical" Gaussian and the remaining parameter to be fit is the
sky background.   
SoDOPHOT never attempts a fit with the 
galaxy model PSF that may be familiar to users of DOPHOT.

The PSF fit parameters from the fiducial stars are 
averaged to determine a PSF model for the new image and the fit magnitudes
are averaged (after removing ``outliers'', i.e. variable stars) 
to find the magnitude offset for
the new image.  
The fit positions of the fiducials are used to determine an accurate
transformation between the template and image coordinates using a general
linear transformation. With the new coordinate transformation, 
a magnitude offset from the template and a new PSF, we
have enough information to construct a model of all the stars in the new
image. The next step is to subtract these model images and then
search the subtracted image for high pixels which are subject to a cosmic
ray test. After the pixels determined to be cosmic rays are removed
from the reduction, the coordinate transformation, PSF and magnitude offset
parameters are refined with a new fit of the fiducial stars.

Next, the entire star list is run through (in order of template brightness
and with the stellar positions fixed)
with each star being added back to the subtracted image and subject
to a two-parameter fit to determine its magnitude
and sky background.  
SoDOPHOT will simultaneously fit pairs of stars
if they are adjacent (known as ``splits''), but no more than two stars at once.
This is the step that generates the photometry for the vast majority of the
stars. After this, there are two more steps designed to improve the
photometry in regions where a significant variation is detected.  First
the subtracted image is searched for high pixel values, and the high
pixels are determined to be either noise spikes, cosmic rays, or new stars.
This procedure ensures that any moderately bright new stars in the image
will be detected. Finally, stars which showed a significant variation
from their template magnitudes and their neighbor are subjected to an
additional round or two of fitting. This reduces the possibility that some
of the flux from the variation of one star will end up being associated
with a close neighbor instead.

Only stars detected in the templates are included in the MACHO database.
Routine SoDOPHOT reductions are then passed through a
filter to search for new microlensing events in real-time.  
We refer to microlensing events detected in real-time as an
``alert events.''  Alert events may be monitored by follow-up networks 
searching for exotic microlensing phenomena.  
A separate database of new objects detected during routine SoDOPHOT 
reductions is also maintained (the SodAlert files).
These files are immediately compressed and exported to a mass store device;
they are not analysed.

A SoDOPHOT instrumental magnitude is the 2.5 times the
base-10 logarithm of the integrated number of electrons in the fit 
to the analytical PSF, 
divided by 100.  No aperture correction is calculated.
SoDOPHOT keeps track of the quality of its photometry with an integer
flag (i-type), which is returned for every measurement.  In Table~1, we summarize
the different integer flags and their meanings. 
SoDOPHOT photometric measurements
flagged as ``unconverged'' or ``obliterated'' are generally regarded
with caution.
These data flags (and others)
are found in the full database and are not necessarily public.  Table~1
illustrates one set of data flags that may be used for quality control on
released MACHO data.

\subsection{Templates}

Images acquired during
the first few months of the experiment were selected
for good seeing and low sky brightness to create
one-time master starlists for each field.   The most important criteria 
for selecting template images were
the depth of the photometry, i.e. the limiting magnitude
and the total number of
stars detected in each image.  
It was not necessary that the template images were obtained 
in photometric conditions.

Templates for LMC and SMC fields were constructed from two observations
to minimize the loss of sky coverage due to the inoperative amplifier (0.0).
Therefore, template photometry is derived from 
3/4 of an image taken from one side of the pier, and 1/4 from the
opposite side of pier.  Templates constructed from CCDs 1, 2, \& 3 (red)
and 4, 5, \& 6 (blue) from the West of pier and CCDs 2 (red) and 5 (blue) from
the East of pier are designated ``West of pier style templates.''
Templates constructed from 3/4 of an East of pier
image and 1/4 West of pier image are designated ``East of pier
style templates.''  
The situation is different for the MACHO bulge data where
all templates are constructed from 
West of pier images only.  
Stars positioned on amplifier 0.0 in the bulge have only blue
photometry.  Templates for the Bulge are designated
``Bulge style templates.''  Bulge style templates
are most similar to West of pier
style templates, except that CCDs 0 and 7 are employed.

Two copies of the template photometry are stored
for use by the MACHO data reduction pipeline (Axelrod et al.~1998)
for LMC and SMC observations.  These two copies
of the master photometry lists are for
East and West of pier observations, and are nearly identical.
However,
the blue photometry for the opposite side of pier from
which the photometry was derived is {\it modified} to approximately
account for the response of the different CCDs and for a focal plane 
position-dependent color gradient (attributed to the dichroic).
This effect is known as ``blue jitter.''  
The modification of the opposite-pier side blue template photometry
is designated the ``blue jitter correction.''
The template photometry without blue jitter 
corrections was used for the database calibration described in the
following sections.

\section{ Calibration of the LMC Top-22 Fields }

\subsection{ The Calibration Algorithm }

It is useful to introduce the adopted calibration
equations and coefficients,
and then review their derivation in greater detail.
The adopted transformation of MACHO instrumental photometry 
to Kron-Cousins $V$ and $R$ uses four coefficents
for each passband: a zero-point, a color
coefficient, a color airmass coefficient (where the airmass
of the template observation is employed), and a chunk offset.
Some stars do not have two-color photometry and so transformations are
only approximate\footnote{In the
LMC top-22 fields, 8468104 out of 9012240 stars (actually objects) have two-color
photometry.}.   The transformation equations have the form,
\begin{equation}
V = V_{M,t} + a0 + \left(a1 + 0.022 \ X_t\right) \ \left(V_{M,t} - R_{M,t}\right) 
+ co + 2.5\log(ET)
\end{equation}
\begin{equation}
R = R_{M,t} + b0 + \left(b1 + 0.004\ X_t \right) \ \left(V_{M,t} - R_{M,t} \right) 
+ co + 2.5\log(ET)
\end{equation}
where $V$ and $R$ without subscripts indicate calibrated
magnitudes on the Kron-Cousins system.  We designate raw MACHO 
magnitudes with the subscript ``M''.  The subscript ``t'' indicates a
template magnitude.  The symbol $X_t$ represents airmass of the template
observation.  
The symbol ``co'' stands for chunk offset.  The standard exposure time 
correction is explicit in equations (1) \& (2);
it is 2.5 times the base-10 logarithm of the exposure time ($ET$) in seconds.
An exposure of 300 sec is used for observations of all LMC fields,
150 sec for Bulge fields, and 600 sec for SMC fields.  

The calibration coefficients are identified as follows.  The zero-point coefficients
in the red and blue are $a0$ and $b0$, respectively.  These coefficients are
common to all stars in any of the 16 chunks on one Loral CCD in a MACHO field.
The zero-points implicitly account for the airmass of the template observation, 
a global aperture correction (mostly seeing dependent), and
also for the possible presence of clouds during the template observation
(non-photometric conditions).  
The color coefficients in the blue and red
are $a1$ and $b1$, respectively.  They correct for the color response
of each Loral CCD at an airmass of zero.  The color airmass
coefficients (0.022 and 0.004) are applicable to all stars.
The chunk offset is
a psuedo-aperture correction relative to the central-most corner
chunk on a given CCD.  It is unique for every field and chunk.

Calibrating MACHO instrumental photometry 
requires {\bf (1) field} and {\bf (2) red West of pier chunk}.
For any object in the database, the field
is known.
It is the first number in the standard
three-integer MACHO database 
identification number~({\it field.tile.sequence}).   
The red West of pier chunk is also known, but is not explicit in the
identification number.
The field number yields the style template
(the layout of the focal plane), and the airmasses of the 
template observations.  The red West of pier chunk specifies 
location in the focal plane, which then uniquely specifies the
zero-points, color coefficients, template airmass,
and chunk offset.

\subsection{ CTIO Observations }

Observations 
were obtained 
in two week-long runs 
on the  Cerro Tololo Inter-American Observatory (CTIO) 0.9-m 
telescope in December of 1994 and 1995
for the purpose of calibrating the MACHO database.  We used the standard
Tek 2048$\times$2048 CCD and $BVRI$ filter set
to make 3 to 4 observations of the center of each MACHO field
using both $V$ and $R$ filters.  Observations of each field were obtained
at a wide range of airmasses and on a minimum of 3 different nights.
The LMC top-22 fields were given priority for observing
when conditions were believed to be photometric.
Typically, we observed 50 secondary standard
stars of Landolt (1992) and Graham (1982) each night at several airmasses.
The secondary standards spanned a range of magnitudes and colors:
$V \sim$ 12 to 18 mag, and $(V-R) \approx -0.1$ to 1.2 mag.
All photometry was performed with
Allstar~II and Daophot~II (Stetson 1987, 1990).  
Transformation solutions for the CTIO instrumental photometry to the Kron-Cousins
standard system were derived for each night using the Landolt and Graham standard
star observations.   We employed a zero-point, color, and mean airmass coefficient
(for each week-long run) in the solutions.  Residuals of these solutions showed
typical standard deviations of 2-3\%, and 4\% on the worst night.
The zero-points derived for the nightly transformation
solutions varied significantly when the airmass and color coefficients
were fixed to run-averaged values, possibly indicating
that the nightly average transparency had changed.

In some cases, the CTIO photometry 
(after applying the nightly transformations
to $V$ and $R$ and aperture corrections) showed $\sim$10\% zero-point variations
from night to night, which may indicate that some of our observations were obtained 
in non-photometric conditions.  However, our aperture corrections
may also contribute to these apparent zero-point variations.  
The large number of images with very few
bright and isolated stars lead us 
to calculate aperture corrections in the following manner.
First, neighbors were subtracted
from around several hundred of the brightest
stars distributed evenly across each images.  For each of these stars, 
standard curves-of-growth
and statistics characterizing each curve were calculated.  
After some experimentation, we opted to
calculate single aperture correction for each observation, ignoring 
any possible CCD position-dependence.  In some cases,
poor subtraction of the neighbors lead to mis-estimates of the
sky and thus inaccurate aperture corrections. 
In order to compensate for this effect, we adopted a rather small 
aperture (radius = 6 pix = 2.4 arcsec) as a measure of the
total flux, which was selected after careful examination
of the Landolt and Graham standard star observations.  

Some repeat observations of the same fields on the same night 
suggested photometric conditions.  Offsets were calculated to shift the 
non-photometric CTIO photometry to these nights and the photometry was averaged.  
Some of the averaged and single observation
CTIO photometry were compared, and no significant
or systematic changes in the colors were found.  It is unlikely that
the MACHO calibration color coefficients (or relative zero-points) were affected
by this choice to average the CTIO photometry. 
In some cases, the averaging appeared to yield tighter sequences 
in the CMDs. 
However, the averaging did not improve 
the precision of the derived calibration coefficients.  
We required that
each star was detected in all CTIO observations for each field,
which yielded approximately 100,000
calibration stars in our final lists.

\subsection{ Comparison of CTIO and MACHO Photometry }

The CTIO tertiary standards are located at the center
of each of the LMC top-22 fields.  The field of view
on CTIO 0.9-m telescope with the
Tek CCD approximately covers the four central-most chunks
of a MACHO field and thus one chunk from each
of the four Loral CCDs (i.e. red West of pier chunks 3, 19, 31, and 51;
see Figure~2).  These chunks are designated the ``zero-point chunks.''
MACHO template photometry without blue jitter corrections
was assembled for the zero-point chunks.
Template coordinates were transformed to a globally consistent
orientation.  The CTIO photometry lists for each field were 
split into four quadrants, and the coordinates were shifted and scaled 
to the MACHO
zero-point chunk photometry lists.
The starlists were then matched using the method of
similar triangles (Groth 1986).   Automated matching of the
starlists was quite difficult
due to the crowded fields and general lack of very bright
reference stars.
In particularly difficult cases, the starlists were 
split again and the procedure was repeated on all possible combinations
of sublists until a satisfactory coordinate transformation was found.

Once the matched photometry lists were assembled, 
we obtained solutions for a variety of different photometric
transformations 
with standard multivariate minimization techniques.  Specifically, we obtained
trial solutions which included non-linearity coefficients, quadratic color terms,
and color airmass terms.   However, these
CTIO data did not satisfactorily
constrain the higher order coefficients or warrant
such complex transformation solutions. 
In additional experiments, we
used different magnitude cuts and
eliminated stars based on their fit to a constant brightness one-year
MACHO instrumental lightcurve.  In the final solutions, no stars were eliminated
for their variability.  

Our final (linear) regressions yielded 88
zero-points and color coefficients.   
We then performed
linear regressions of the color coefficients with the template airmass data  
to derive the best fit CCD color coefficients at an airmass of zero ($a1$ and $b1$) 
for the adopted color airmass coefficients (0.022 and 0.004).  The color airmass
coefficients were also indicated by these regressions, 
but were somewhat poorly
constrained.  Therefore, we fixed their values (see also \S6.2 of this paper)
and then derived the
$a1$ and $b1$ coefficients. 
The uncertainties in the derived values of the $a1$ and $b1$ coefficients 
are estimated to be $\sim$0.005;
these coefficients are listed in Table~2.
The color coefficients for CCDs 0 and 7 are adopted.
They are never used for LMC or SMC calibration, only for the bulge.

Once the color coefficients were determined and fixed,
the CTIO data were used to derive
a single zero-point and three relative CCD offsets 
(for each color) for each of the LMC top-22 fields.  
Trials with different data subsets 
indicated a typical uncertainty for zero-points and offsets of order
$\sim$0.03 mag.   
In Figure~3,
we present a CMD showing
the CTIO data used to calibrate four zero-point chunks in MACHO field 13.
We also show the difference between the calibrated CTIO and MACHO mags
as a function of $V$ and $R$.  Approximately 900
stars are plotted for each zero-point chunk.  
This figure illustrates the typical magnitude and color range of
the CTIO tertiary standard stars,
and the typical dispersion
of these data about the derived zero-points.

\subsection{ Chunk Offsets }

Chunk offsets are aperture corrections relative
to the zero-point chunk photometry for each CCD.
The chunk offsets for the 
zero-point chunks are set equal to zero by definition.
Aperture photometry of several thousand
bright stars in each MACHO template image was obtained with Daophot.
It was necessary to re-reduce the template images 
with the appropriate flats and gain tables, as these have changed
during the course of the experiment and the original reduced template
images were not saved.  
It was assumed that the Daophot aperture photometry, in a relative
sense, uniformly measured the flux for these bright stars across
the entire image.  
MACHO instrumental template photometry was then assembled, 
coordinates were transformed to the original image/pixel system,
and stars were matched to the Daophot aperture photometry lists.
Chunk offsets were derived by comparison of the Daophot
aperture photometry with the SoDOPHOT template photometry.

Chunk offsets were derived for both the
red and blue template images.  
Comparisons of photometry in field-overlap regions calibrated
with the blue chunk offsets 
showed worse agreement than comparisons of photometry
calibrated with no chunk offsets.  Calibration
trials applying chunk offsets derived from the red image data 
to both the red and blue photometry
showed the best agreement in the field-overlap comparisons.
The chunk offsets likely reflect the ability of the SoDOPHOT analytic
PSF to fit the real instrumental PSF, which may change
across the focal plane.  In this case, the red image data appears
to yield a more accurate measurement of the effect, for the technique
we have used to calculate the chunk offsets.
The chunk offsets are correlated with position in the focal
plane, which is consistent with our explanation of their origin.
In Figure~4 we plot the mean of the LMC top-22 field chunk offsets
for each of the 64 chunks (red West of pier chunk number
is in the lower right number of each box). 
The typical standard deviation for the mean chunk offsets
is 0.02 mag.

\subsection{ Field-Overlap Comparison }

Many of the MACHO fields in the LMC top-22 bar fields overlap
the same region of sky\footnote{There is no complete census of field-overlap
stars in the MACHO database, although it
is estimated to be 6.5$\%$ of the total number of stars
in the 22 bar fields.
For a map of these fields, see Fig.~1 of Alcock et al.~(1997).}.
The uncertainties of MACHO astrometry (typically an arcsecond) and the 
crowded nature of the bar fields complicates the identification of
field-overlap stars.  Probable pairs of field-overlap chunks
were found via inspection of a map of our fields.  Files
of photometry with amplifier coordinates transformed
to a globally consistent orientation were matched using
the method of similar triangles (Groth 1986).
In this manner, approximately
360,000 stars in 150 chunk pairs were identified in field-overlap regions.
These overlap regions allow us to check the precision of the photometric
calibration for 21 of the top-22 LMC bar fields (field 47 is isolated).
It was typical to measure the median offset between calibrated $V$ and $R$
magnitudes for two chunks in field-overlap regions with a precision 
of $\sim$0.03 mag.

At this point in the calibration campaign, we attempted to globally minimize
the zero-points and offsets.  
This was a somewhat subjective procedure which required a high
level of human interaction.  We began with a handful of fields as
calibration ``anchor points'' and then 
allowed the other field's zero-points to vary.
Zero-points were adjusted to minimize the 
offsets in field-overlap regions.  In some cases,
we also adjusted the CCD offsets (the zero-point offsets for the different
CCDs relative to the single 
field zero-point determined from the CTIO data).  However, these 
were never changed by more than the estimated 1$\sigma$ uncertainty
of their measured values ($\sim$0.03 mag).  We additionally made
a handful of ``reality checks''
(i.e. comparisons with published photometric standard sequences in our
fields) through-out this global minimization procedure.  Finally, the entire
global minimization procedure was repeated several times.  Each time we chose different
anchor points and varied the sequential order of the field-by-field comparisons.
In this manner, we endeavoured to minimize possible systematic errors 
introduced by this procedure.

{\it The MACHO calibrations have an internal precision
of $\sigma_{V}$ = 0.021 mag,
$\sigma_{R}$ = 0.019 mag, and $\sigma_{V-R}$ = 0.028 mag} for
$\sim$9 million stars distributed over $\sim$10 square degrees of sky.  
This is illustrated in Figure~5 where we plot $\sim$20,000 stars in 150 chunks
with $V < 18$ mag.  We plot the offset in $V$, $R$, and $(V-R)$ versus
magnitude or color in the top, middle, and bottom panels respectively.
The standard deviations
given above (and labeled in Figure~5)
are calculated from the median offsets determined for each of 
the 150 field-overlap chunk pairs.

\section{ Comparisons with Other Photometry  }

\subsection{ Ground-based Data }

It is customary to compare newly calibrated photometry with 
previously published data.  
We have arbitrarily
selected $\sim$200 stars representing
photographic, photoelectric, and CCD data from a dozen different authors. 
This may be a representative sample.  We restrict this comparison 
to our $V$ photometry, because $R$ photometry is less common.  
In \S4.2, we make comparisons with our $V$ and $R$ calibrated photometry,
which allows the reader a more comprehensive assessment of the data.

We begin 
with a comparison of nine period-folded $V$ band lightcurves for an
arbitrary sample of RR Lyrae and classical Cepheid variables.
In Figure 6, from top to bottom then left to
right we plot the classical Cepheids: HV900, HV905, HV2510, HV2352, and HV2324,
and the RR Lyrae near the cluster NGC~1835: GR-6, GR-14, GR-16, and Walker-V26.
The MACHO data consists of $\sim$1000 measurements; they are plotted as dots.
Error bars are omitted for clarity.  
The comparison data are plotted as filled circles and are assembled from
Martin and Warren (HV900, HV2323, HV2352; 1979), Sebo and Wood (HV905; 1995),
Martin (HV2510; 1981) and from Walker (the RR Lyrae; 1993).  For the
RR Lyrae, the finding charts from Graham and Ruiz (1974) were also employed.  
We note that the photometry of Moffett et al.~(1998; not plotted) for HV900 
also shows satisfactory agreement with the MACHO data.  
All of the RR Lyrae are located in the
same MACHO field and chunk, thus the $\sim$0.2 mag differences 
in brightness from star to star may not
be attributed to a calibration error.  These RR Lyraes illustrate 
the difficulty obtaining
accurate photometry in crowded fields at this brightness.  

In Figure 7, we compare our $V$ photometry with various other data.
We plot
$\delta V$(MACHO$-$Other) versus $V$ mag.  We designate different author's data
with different symbols as follows.
Asterisks are the Cepheids
and RR Lyrae from Figure~6, 
filled circles are Walker's (1993) standard star sequence near
NGC~1835, open triangles are standard sequence 
of Cowley et al.~(1990) near Cal-87
(finding chart found in Pakull et al.~1988), 
filled triangles are data from Flower et al. (1982) near NGC~2058/2065, 
open squares are stars
near NGC~1847 from Nelson and Hodge (1983), and the open circles are
photometry assembled 
from the classical LMC bar photometry paper by Tifft and Snell (1971).
The median offset between MACHO and all of the other data is
$\delta V = -0.035$ mag, which is indicated with a dashed line. 
Differences among the various authors likely represent
systematic calibration errors.   

\subsection{ HST Data }

We have obtained {\it Hubble Space Telescope} (HST) Wide Field Planetary
Camera (WFPC2) image data with the F555W and F675W filters 
for three fields in the LMC top-22.  These are located
in the field-overlap region of MACHO fields 2 \& 79, in field 13, and 
in field 11.
The reader is referred to Alcock et al.~(1997) for a map 
and sky coordinates of these fields.
For each field, we obtained ``shorts'' (3-4 $\times$ 30 sec exposures) and ``longs''
(3-4 $\times$ 400-500 sec exposures), in both filters.

Complete details of the HST data reduction
and analysis will be presented elsewhere
(Nelson et al.~1999).  Briefly, the images
were co-added and geometrically corrected with the 
aid of the stand-alone Drizzle package (Fruchter \& Hook 1998).
We performed aperture photometry 
with the centroids accurately determined via PSF fitting using
Allstar~II/Allframe and Daophot~II (Stetson 1987, 1990).
Stars were matched between filters
for the longs and shorts separately, and for each field observed.
Only stars identified in both colors were kept in the final photometry lists.
The median aperture correction was calculated from
$\sim$50 bright and isolated stars
per WF chip, per short or long exposure, and per filter.   
The photometry lists were calibrated
according to Holtzmann et al.~(1995) using the 
coefficients in their Table~7 and the bay~4 gain ratios.
The short and long photometry lists were then combined,
including all stars found in either list, and adopting the
long exposure magnitudes for stars in common to both lists.  
We made no correction for the WFPC non-linearity. 

In Figures 8, 9, and 10 we compare approximately 120 stars
identified in both the MACHO and HST photometry data.  For each
field we show the difference in magnitude, $\delta V$ and $\delta R$
(MACHO $-$ HST), as a function 
of $V$ or $R$ (top two panels) and $\delta(V-R)$ versus
$(V-R)$ in the bottom panel.  The open triangles are data from the WF2,
the open circles are from WF3, and the open squares are from WF4. 
The median offset of the data is labeled in each panel.  Typical dispersion
about these median offsets is 0.02 mag.  A direct comparison of ground-based
and HST data is complicated by the radically different resolutions.  We have
simply added the flux for all HST stars inside a 1'' radius of the star 
identified as the match to the MACHO object.  
Various trials with different
``artificial blending'' schemes indicate that the precision of this comparison 
is $\sim$0.05 mag.  It is difficult 
to identify the magnitude and angular separation (among pairs or groups
of stars) where the faint HST stars
``become sky'' and would no longer be counted 
in the ground-based SoDOPHOT measurements. 
Resolution of this issue
is beyond the scope of this work.

The MACHO and HST photometry comparisons in
Figures 8-10 show agreement at the $\sim$5\% level or better.   
We note that there are no systematic
differences between the photometry derived for the three WF CCDs and the MACHO
photometry (in each case, derived from a single CCD image), in three separate
comparisons.  Our fields are particularly useful for this comparison, because
they are fairly uncrowded (for HST) and the comparison stars are distributed
throughout each WF CCD, in each field.  
The apparent consistency of the calibrated WF photometry supports the procedure
we have used for calculating the aperture corrections and also 
supports the calibration formulae for WFPC2
given by Holtzmann et al.~(1995).  Given the
$\sim$5\% uncertainty associated with 
blending the HST data to match the MACHO data, 
these comparisons indicate that
$R$ agrees quite well, and that the calibrated colors 
are offset by 0.04 mag ($(V-R)_{HST} \approx
(V-R)_{MACHO} + 0.04$).

In Figure 11, we present two
side-by-side CMDs showing the 120
MACHO objects and the ``un-blended'' HST photometry from the comparison above.  
We plot the MACHO objects in the left panel and 
the HST stars in the right panel.
Without making any
corrections for completeness in the HST data, this naive comparison
yields a star to object ratio $S/O$ = 229/120 = 1.91.  This value is sensitive to
the adopted match radius (1''), but may reasonably be
considered a lower
limit to the average star to object ratio in these three fields.
We will return to this in \S7.

\section{ Other MACHO Fields }

\subsection{ Bulge Field 119 and SMC Field 207 }

We have calculated calibration zero-points for field
119 (Baade's Window) in the Galactic bulge and field 207 (near the cluster
NGC~330) in the SMC by matching published photometry to MACHO data.

MACHO field 119 is centered on Baade's Window.  Calibrated
photometry of $\sim$70 stars from Cook (1986) and 7 standard stars from Walker 
and Mack (1986) in the $V$ and $I$ passbands were converted to $V$ and $R$ using
$(V-R) = 0.50(V-I)$ (e.g.~Landolt 1992).
These stars were then identified in the MACHO photometry database.  
The MACHO photometry was corrected for color and airmass according to 
equations (1) \& (2) of this paper.  Chunk offsets were assumed
to be zero everywhere.  In Figure~12,
we plot the difference between the standard and color corrected
instrumental MACHO magnitudes 
as a function of $V$ mag.   The same plot is shown in the bottom panel
for the $R$ mags.
The median values of these magnitude differences are our
adopted solutions for $a0$ and $b0$, and are plotted as solid lines
in the top and bottom panels respectively.  
The zero-points are 
$a0 = 18.259$ and $b0 = 17.972$, with a standard deviation of 
$\sim 0.025$ mag.

The well-studied SMC cluster NGC~330 is located in MACHO field 207.
We have used photometry from Vallenari
et al.~(1994) near NGC~330 to calibrate our SMC photometry.  These
data are in $B$ and $V$ and we have converted them to $V$ and $R$
using $(V-R) = 0.56(B-V)$.  
The MACHO instrumental photometry was corrected for the color 
and airmass according to equations (1) \& (2).
We plot the data used
for the field 207 zero-point solutions in 
Figure~13.  We find
$a0 = 17.788$ and $b0 = 17.584$, with a standard deviation of 
$\sim$0.07 mag; these are indicated with solid lines in Figure~13.

\subsection{ Status of Calibration for All Other MACHO Fields }

In Figure 14 we plot the LMC top-22 zero-points 
($a0$ top panel, $b0$ middle panel, and $a0 - b0$ in the bottom panel)
as a function of template airmass ($X_t$)
using small open circle symbols.  The single
bulge zero-point is plotted with a filled circle symbol, and the single
SMC zero-point is plotted with a filled triangle symbol.  
Linear regressions of the LMC top-22
zero-point data are indicated with dashed lines in each panel.  We will use
these to predict approximate zero-points for all other MACHO fields in the
LMC, SMC, and bulge.  The derived regressions are,
\begin{equation}
a0 = 18.410   -  0.279 X_t
\end{equation}
\begin{equation}
b0 = 18.087   -  0.222 X_t
\end{equation}
The uncertainty in the zero-points are $\pm 0.06$ and the uncertainty
in the slopes are $\pm 0.04$.
The dispersion of the data about each of these regressions
is $\sim$0.10 mag.  
The datapoints lying below the regressions are likely
due to non-photometric conditions (i.e.~clouds), while the points above
tend to have good seeing (this represents a by-field aperture correction,
relative to the typical observation).
At this time,
the calibration zero-point uncertainty for all fields
in the MACHO database which have not been explicitly calibrated
in this paper is 0.10 mag.  
The uncertainty in color is 0.04 mag, which is estimated
from the dispersion about the $a0 - b0$ regression in the bottom panel
of Figure~14.

\section{ Lightcurve Calibration }

\subsection{ Blue Jitter }

MACHO instrumental lightcurves in the blue show a systematic 
source of scatter which is directly attributable to the 
responses of the different Loral CCDs and a secondary focal plane
position-dependent effect likely due to the dichroic.  This is 
blue jitter.  Because observations of LMC and SMC fields are made from
both sides of the pier, stars will alternatively land on the CCDs rotated
180 degrees from each other in the focal plane (i.e.~a star will land in
CCDs 4~\&~6, or 5~\&~7 depending on pier side).  
The opposite-pier side template photometry 
is modified (i.e.~the
template photometry is ``jittered'') so that PSF fitting in SoDOPHOT
will converge quickly.
This has the effect of
{\it maintaining} the systematic differences in blue photometry derived from different
pier side observations in the resulting instrumental lightcurves.
It is possible to ``de-jitter'' lightcurves by applying the inverse of
the template photometry blue jitter correction.  
The de-jitter algorithms have the form,
\begin{equation}
V_{M,t} = V_{M,e} + BJ_e \ \left[ \left( V_{M,e}-R_{M,e} \right) - BJ_o  \right]
\end{equation}
\begin{equation}
V_{M,t} = V_{M,w} + BJ_w \ \left[ \left( V_{M,w}-R_{M,w} \right) - BJ_o \right]
\end{equation}
where the subscripts ``e'' and ``w'' stand for East and West respectively.
The subscript ``t'' stands for template magnitude; thus $V_{M,t}$ 
calculated from equations (5) \& (6) is an
appropriate magnitude for input into equations (1) \& (2).
De-jitter corrections require
an instrumental color for each measurement.  
They also require the three coefficients
$BJ_e$, $BJ_w$, and $BJ_o$.  

The coefficients $BJ_e$ and $BJ_w$ depend on 
(red West of pier) chunk and field.  The field
gives the style template, and the chunk specifies location in the focal
plane.  The coefficients $BJ_e$ and $BJ_w$ are unique for each chunk
and applicable to all fields.  However, the sign of the coefficient 
flips depending on the style template, and for every chunk,
either $BJ_e$ or $BJ_w$ is set equal to zero.  The $BJ_{w}$ coefficients
in Figure~15 are for blue jitter corrections in the imaginary case 
of a West of pier observation made of a field with template photometry derived from
an entirely East of pier observation.  This illustrates the focal plane dependence.

The coefficient 
$BJ_o$ is unique to every field and chunk, and is calculated from
the mean color of the PSF stars in that chunk.  
$BJ_o$ is simply the color in each chunk
where no blue jitter correction is necessary\footnote{The de-jitter correction
applied to lightcurves for the year-one and two-year LMC microlensing analyses
may be approximately recovered by setting $BJ_o$ equal to zero everywhere.}.
The calibration between $BJ_o$ and the mean color of the PSF stars in
a chunk was derived by minimizing the difference of mean blue magnitudes
calculated using only East and West of pier data in the four-year lightcurves
of $\sim$400 constant brightness stars (per chunk),
for 224 different chunks in 5 different LMC fields (9, 10, 18, 19, 82).
These data are shown in Figure 16 along with our adopted calibration
(shown as a solid line), which follows: for mean PSF colors 
$(V-R)_{\overline{PSF}}$ $<$ $0.35$ mag the coefficient $BJ_o$ = $-0.05$ and
for $(V-R)_{\overline{PSF}}$ $>$ $0.35$ mag the coefficient 
$BJ_o = -0.4$ + $(V-R)_{\overline{PSF}}$.
We estimate that de-jittering will be
accurate to $\pm 0.02$ mag with this procedure.  
The mean color of the PSF stars has
been calculated for every chunk in the MACHO database.

\subsection{ Color Airmass }

MACHO instrumental lightcurves may show 
systematic changes in brightness which are correlated with the airmass
of individual observations.  Due to the limited observing season 
and the high priority set for
observing the LMC, 
these fields are observed at progressively higher 
airmasses as the season progresses.  This may result in slow ``seasonal rolls'' 
($\sim$1 year periods) in some instrumental 
lightcurves.
We present here an algorithm for an airmass and 
color-dependent lightcurve correction. 
This correction is
{\it not implemented} in any existing MACHO calibration code, but
may potentially be useful for correcting small numbers of astrophysically
interesting lightcurves.

The color airmass corrections should be made on East and
West of pier lightcurve data separately before making blue jitter 
corrections. We show the form of
the correction for East of pier data, maintaining the use of the subscript
``e'' to designate these measurements.  The correction is equally applicable
to the West of pier data.  We designate the $raw$ 
magnitudes with a prime symbol.  The color airmass corrected
data does not have the prime, and would be appropriate for substitution
into equations (5) \& (6) for blue jitter corrections.  The color airmass
corrections have the form,
\begin{equation}
V_{M,e} = V_{M,e}^{'} + 0.033 \ \left(X - X_t \right) \ 
\left[ \left(V_{M,e}^{'}-R_{M,e}^{'} \right) - CA_o \right]
\end{equation}
\begin{equation}
R_{M,e} = R_{M,e}^{'} + 0.006 \ \left(X - X_t\right) \ 
\left[ \left(V_{M,e}^{'}-R_{M,e}^{'}\right) - CA_o \right]
\end{equation}
where $X$ is the airmass of each observation, $X_t$ is the airmass of the
template observation, and $CA_o$ is the color for which no 
airmass-correlated changes in brightness are observed.  
The ratio of the color airmass coefficents in equations (1) \& (2)
to the color airmass
coefficients given in equations (7) \& (8) are $V$: 0.022/0.033 and $R$:
0.004/0.006.  Both ratios are $\sim$0.66, which is approximately
$(V - R)$/$(V_{M} - R_{M})$.  These color airmass coefficients
are derived from MACHO instrumental lightcurves, and support the values
used in the calibration formulae, i.e~equations~(1) \& (2).

The $CA_o$ coefficient varies from chunk to chunk, although no correlation
was found with the mean color of the PSF stars, as was the case for
the blue jitter $BJ_o$ coefficient.  
Therefore, in order to make the color airmass lightcurve correction, 
$CA_o$ must be derived
using separate knowlege of the shape of a lightcurve.  For example,
one could solve for $CA_o$ by minimizing the scatter in the period-folded lightcurve 
of a Cepheid variable star.  An alternative way to solve for $CA_o$ would
be to minimize the scatter in several constant brightness stars in the same chunk
as the star of interest.  
In this case, it is recommended
that several nearby stars are used in the solution, and that they have a wide range
in colors, ideally bracketing the color of the star of interest.  
It is recommended
to inspect scatter plots of $V_{M,e}$-$X$ and $V_{M,w}$-$X$ (separately, as blue 
jitter may mask the effect) in order to estimate the
degree to which any particular lightcurve is affected.

\subsection{ Other Systematic Lightcurve Effects }

We offer a few cautionary remarks to potential
users of MACHO data.  
{\bf(1)} Weather permitting, the MACHO Project will observe every
night of the year.  This includes nights when the seeing
(the FWHM of the stellar PSF) approaches $\sim$7 arcsec.  
In our typically crowded fields,
poor seeing can lead to inaccurate photometry, despite
the small photometric uncertainties that may be reported by SoDOPHOT.
Inspection of scatter plots of $V_{M,w}$- and $V_{M,e}$-$Seeing$ are
useful diagnostics for this source of systematic lightcurve scatter.  
This so-called ``seeing variability'' can be quite large (easily 
a few tenths of a magnitude),
particularly for very crowded stars or stars nearby to 
regions of irregular, bright nebulosity.  It is not recommended to
globally decorrelate lightcurves with seeing.
{\bf(2)} The catalogs of CCD defects
polluting the MACHO focal plane are not perfect.  Uncataloged
CCD defects will cause spurious photometric measurements, which are
not necessarily reflected in the photometric uncertainty or integer
flags reported by SoDOPHOT.  In this case, 
inspection of the image data is 
very useful.  {\bf(3)} In some cases, 
lightcurves will exhibit
additional scatter over that expected from the uncertainties
of individual measurements
which is not attibutable to any of the aforementioned
effects.  This may be due to the inclusion of
variable stars in the PSF fiducial lists used by SoDOPHOT.  In this
case, other lightcurves in the same chunk may also be affected.

\section{ The Efficiency CMD }

The efficiency calculation will be reported in complete detail elsewhere
(Alcock et al.~1999).  
Briefly, the calculation is a massive series of artificial
star tests with raw image data spanning four years of observations followed by
Monte-Carlo experiments to detect fake microlensing events in accurately
simulated artifical datasets.  
The efficiency calculation returns our sensitivity to detect microlensing
in stars (not ``objects'', see \S1).  We calculate the star to object
ratio in the MACHO data by comparing with the HST data.
We assume that there are no objects in the
high resolution image data from the HST, only
real stars\footnote{Observational data and further discussion
of binary source stars (and lenses) will be presented in a forthcoming
paper.} .

The efficiency CMD represents
real LMC stars and contains
accurate information on their surface density as a function
of color and magnitude.  For convenience, we
use an area normalization
of 0.52 square degrees (one MACHO field).  
The limiting magnitude
of the efficiency CMD is set by the faintest LMC star for which we may 
realistically detect microlensing.  
We adopt $V$ = 25 mag, indicated by preliminary results
from the efficiency Monte-Carlos. 
The efficiency
CMD plays two roles in the efficiency calculation: 
(1) it is used to seed the artificial star tests with stars
drawn from a realistic distribution in color and brightness, and (2) the derived
efficiency (for stars) must be integrated over the efficiency CMD to calculate
the efficiency per object in the MACHO database.  
The required accuracy of the efficiency CMD is dictated by the second application.
The efficiency CMD has direct consequences for our
measurement of the LMC microlensing optical depth.

In Figure 17, we compare the three MACHO CMDs for fields
2, 11, and 13 for which we also have HST data
(these are actually log-scaled Hess diagrams).
Fields 2, 11, and 13 contain 354586, 426060, and 344746 objects 
respectively.  We note that field 13 has the faintest limiting magnitude.
Also, with the exception of the varying degree of differential reddening (field 11
is the most affected), and slight differences in the
numbers of upper-main sequence stars, 
these three CMDs are quite similar.

In Figure 18, we compare the MACHO object luminosity functions 
with the HST star luminosity functions (LFs).  
We plot $dN/dV$ as a function of $V$ mag.  The units of $dN/dV$ are $10^4$
stars (or objects) per 0.52 square degree in 0.125 $V$ mag bins.
The MACHO data are shown as solid lines.  The HST data are shown 
as open circles connected by dotted lines.   The typical
error bar for the HST data is 
indicated in the upper right corner.  We compare fields 2, 11, and 13 in the 
top, middle, and bottom panels, respectively.  
The HST data has been scaled to the MACHO data by a factor of 409, which 
is estimated as follows.
The effective area photometered for each HST field is
is $3 \times 747.5 \times 747.5$ pixels.
The plate scale of 0.''1/pix yields a sky area of 4.6 square arcmin per field.
The MACHO plate
scale of 0.''635/pix, and mosaic of $4 \times 2048 \times 2048$ Loral CCDs
yields a total sky area of 1879.1 square arcmin per field.  We scale
the HST data by the ratio of sky areas after making a small (2\%) correction
for completeness in the HST data (see below).
Figure 18 shows that the MACHO photometry is incomplete
at the brightness of the red horizontal branch clump ($V \sim 19.3$ mag) 
in fields 2 and 11.
In constrast, field 13 appears complete to $V \sim 21$ mag.

In Figure 19, we compare the HST LFs for the three fields.  
We plot the logarithm of $dN/dV$ (same units
as in Figure 18) as a function of $V$ mag.  Fields 2, 11, and 13
are shown as dotted, short-dash, and long-dash lines respectively.
The sum of these three LFs is shown as a bold line.
Preliminary artificial star tests indicate that
we are $\sim$ 98\% complete to $V \sim 22$ mag.   We have fit a
power-law to the summed LF in the magnitude range $V = 19.5$ to 22 mag,
and extended it to $V$ = 25 mag (the slope of the derived power-law
is $\alpha = 0.415 \pm 0.017$).
The power-law LF is shown as a solid line.  We assume this closely
approximates the true LF.  
Deviations from the
power-law may be dues to incompleteness in the HST data
or reflect a real turn-over.
We will use (below) the ratio the power-law LF to the
summed HST LF to re-scale the number of stars in
the efficiency CMD from $V$ = 22 to 25 mag. 
Considering only stars with $V < 25$, the summed HST LF contains
17006732 stars.  The summed HST LF for $V < 22$ plus the power-law
LF for $22 < V < 25$ contains 55864940 stars.

In order to improve the sampling of the sparsely populated
bright end of the efficiency CMD, we will ``splice'' 
the bright MACHO data to the faint HST data.
We will make the splice
at $V = 18.7$ mag, where brighter than this, all three 
MACHO fields appear to be complete (see Figure 18).
The HST data and the MACHO data for the three fields are
each binned in color from $(V-R)$ = $-0.5$ to 1.5 in bins
of 0.05 mag and in brightness from $V$ = 25 to 15 in bins
of 0.10 mag.  These 2-d histogram data are converted to images
compatibile with IRAF\footnote{The Image Reduction and Analysis
Facility, v2.10.2, operated by the National Optical Astronomy Observatories.}.
Within IRAF, we replace all pixels fainter than $V$ = 18.7 mag in the
MACHO CMD image with zero values, and similarly edit the HST image
pixels values brighter $V$ = 18.7 mag.  We additionally replace 
low density pixels with zero-values in each image.  
This step typically ``removes'' the few pixels containing galaxies, foreground
stars, or bad measurements.  In the HST image, these pixels 
would otherwise be scaled to very
high values in the efficiency CMD.   
In the MACHO image, this step also removes features such as the upper-main
sequence, asymptotic giant branches, supergiants, and
foreground Galactic disk stars.  The regions of the CMD populated by these
features are excluded
in the searches for microlensing anyway.  This editing has a negligible effect
on total number of stars in the efficiency CMD.  

Next, we sum the edited HST and MACHO images and smooth the 
resulting image.
We multiply the spliced and smoothed CMD image by 
a power-law scaling image. 
The resulting efficiency CMD is shown in Figure~20.
Intensity represents the number of stars.  The image is log-scaled;
contours indicate 1.0, 3.5, 4.0, 4.5, 5.0, and 5.5 dex.
The total number of stars represented is 55426384 (per 0.52 square degrees).
The corresponding number of MACHO objects is 1125392,
which yields a star to object ratio $S/O$ = 49.2 to $V$ = 25.  
We find $S/O$ = 1.2, 3.4, and 21.8 to
$V$ = 20, 22, and 24 mag respectively.  These
values are systematically uncertain at the $\sim$8\% level, depending
on the counting of MACHO objects identified only in one color.

It may be possible
to adopt a single efficiency CMD for the entire LMC and then use the efficiency
Monte-Carlos to
sort-out our relative sensitivities to detecting microlensing
in different fields.  However, we may also assess the uncertainty associated 
with the single
efficiency CMD approximation as follows.  
Consider the three HST LFs in Figure 19.   While the overall distributions
are quite similar, 
fields 2, 11, and 13 contain 1335794, 1406142, 595913 stars (scaled to 0.52 square 
degrees) with $V < 22$ mag.  Using the total number of MACHO objects in these
fields, we find $S/O$ = 3.8, 3.3, and 1.7 for fields 2, 11, and 13 respectively,
which yields $<S/O> = 2.9 \pm 1.0$.  The standard deviation indicates the
uncertainty of our single efficiency CMD approximation.  Because of the
power-law LF corrrection 
for $V > 22$ mag, $S/O$ to all fainter magnitudes scales
directly from the $V \sim 21$ mag portion of the HST LFs.

We may extend this analysis to other fields using surface brightness
measurements in the LMC.
First, we measure relative fluxes (arbitrary
units) for 16 MACHO fields from the Bothun and Thompson (1988)
``Parking Lot Camera'' wide-field R-band image of the LMC (the fields and
this image are shown in Fig.~1 of Alcock et al.~1997).  We exclude the
central-most bar fields because the image is saturated here.  We also exclude
two fields near 30 Doradus.  The 16 fields selected span the ``middle
range'' of the 30 fields targeted for the upcoming 
LMC microlensing analysis in terms of total number of objects.
For our three fields
with HST data, we perform a linear regression between the HST-derived $S/O$
ratios and their surface brightnesses.  We then use surface brightness to
predict $S/O$ for the remaining fields.    These data are shown in Figure 21
where we plot $S/O$ as a function of total number of objects per field
(in units of $10^5$ objects).  
Filled triangles are the $S/O$ values calulated from the HST data for
fields 2, 11, and 13.  The filled circles are the fit values from the
surface brightness regression for these three fields.  
The open circles are the 
estimates $S/O$ values for the remaining fields.  
We are primarily concerned with 
the estimated scatter; the standard deviation 
is $\sim$0.80, or a 30\% uncertainty in $S/O$ (also labeled on
Figure 21).

Figure 21 suggests a correlation between
$S/O$ and total number of MACHO objects per field.  
It may be the case that additional parameters (i.e. seeing or sky
in the template images) would improve the correlation.  If so, such
a correlation may be used
to rescale the efficiency CMD and
improve the accuracy of the efficiency calculation,
particularly as a function of position in the LMC.  
However, this analysis awaits further characterization of
our SoDOPHOT-derived photometry and the efficiency Monte-Carlo results.

\section{Summary}

In this paper, we have described MACHO Project photometry
data and calibration of these data.
We have transformed our two-color instrumental 
photometry for $\sim$9 million stars in the LMC top-22 fields (Alcock et al.~1997)
to the Kron-Cousins $V$ and $R$ system
with a precision of $\sigma_V = 0.021$, $\sigma_R = 0.019$,
and $\sigma_{(V-R)} = 0.028$ mag.  The uncertainties associated with
the CTIO photometry aperture corrections and the MACHO transformation
zero-points are the most significant for the overall accuracy of the
photometric calibration.  We estimate $\pm0.018$ for the former,
and $\pm 0.03$ mag for the latter, which
is likely a correlated error in $V$ and $R$.  
Therefore, we estimate the overall accuracy of
these calibrated MACHO photometry to be
$\pm 0.035$ mag in $V$, $R$, and $(V-R)$. 
This appears consistent with comparisons
of calibrated MACHO photometry data to
published photometric sequences and calibrated HST observations.
The accuracy of the calibrated MACHO photometry may be worse for
stars with extreme colors due to our non-standard passbands.
Calibration for all fields 
not including the LMC top-22 (Alcock et al.~1997),
SMC field 207, and Galactic bulge field 119
has a zero-point uncertainty of 0.10 mag and
a color uncertainty of 0.04 mag.  
The calibration presented in this work
supersedes all prior calibration of MACHO photometry in the LMC, SMC,
and Galactic bulge.

The detailed descriptions of the data reduction and photometry
database provided in this work are 
largely intended to guide consumers of released MACHO data.
If potential calibrators 
have assembled MACHO instrumental template
photometry, or otherwise adopted/calculated MACHO instrumental
magnitudes and colors as substitutes for the
original template photometry, the necessary calibration formulae
are given in \S3.1.
We have additionally discussed the calibration of MACHO lightcurves.
If potential calibrators will make detailed analyses
of MACHO lightcurves, they should consider making
corrections for known systematic effects 
such as blue jitter (\S6.1), color-dependent airmass variability (\S6.2),
as well as effects due to seeing and uncatalogued focal
plane defects (\S6.3).
Last in this work, 
we have described the construction of the efficiency CMD, a cornerstone
of the new microlensing detection efficiency calculation,
which will be presented in a forthcoming
MACHO collaboration paper.

\section{Acknowledgements}

David Alves thanks his graduate advisors, Kem Cook
and Robert Becker, for their support.
We thank the skilled support by the technical staff at MSSSO.
Work at LLNL supported by DOE contract W7405-ENG-48.  Work at CfPA 
supported by NSF AST-8809616 and AST-9120005.  Work at MSSSO supported
by the Australian Dept.~of Industry, Technology and Regional Development.
WJS thanks PPARC Advanced Fellowship, KG thanks DOE OJI, Sloan, and
Cottrell awards, CWS thanks Sloan and Seaver Foundations.

\clearpage
\begin{deluxetable}{cl}
\footnotesize{}
\tablewidth11cm
\tablecaption{SODOPHOT Integer Flags\tablenotemark{A}}
\tablenum{1}
\tablehead{ 
\colhead{i-type\tablenotemark{B}}  &    
\colhead{meaning} 
}
\startdata
 1  & bright, unsplit star \nl
 2  & bright, split star   \nl
 3  & faint, unsplit star  \nl
 4  & faint, split star    \nl
 5  & very faint or unconverged unsplit star    \nl
 6  & very faint or unconverged split star      \nl
 7  & too faint or off the image, unsplit star  \nl
 8  & too faint or off the image, split star    \nl
 9  & unconverged in 7 parameter fit  \nl
 10 & large number of pixels missing ($>$ 35\%)  \nl
 11 & cosmic ray  \nl
 12 & galaxy (currently disabled)  \nl
 13 & obliterated star  \nl
\enddata
\tablenotetext{A}{These data flags (and others)
are found in the full MACHO database and are not necessarily public.  This
table lists one set of data flags that may be used for quality control on
released MACHO data.}
\tablenotetext{B}{The i-type 
is modified by a boundary flag: i-type = i-type(as above)
+ $20\times$(template boundary flag) + $40\times$(boundary flag).
The (template boundary flag) = 1 for stars within 10 pixels of a template
chunk boundary and 0 otherwise. The (boundary flag) = 1 for stars within 
10 pixels of a routine reduction chunk boundary, = 2 for stars which fall
off the image altogether, and = 0 otherwise.}
\end{deluxetable}

\clearpage
\begin{deluxetable}{cccc}
\footnotesize{}
\tablewidth11cm
\tablecaption{Calibration Color Coefficients}
\tablenum{2}
\tablehead{
\colhead{Blue CCD No.} & \colhead{$a1$} &
\colhead{Red CCD No.}  & \colhead{$b1$}
}
\startdata
 1 & $-0.1876$ & 6 & 0.1868 \nl
 2 & $-0.2065$ & 5 & 0.1784 \nl
 3 & $-0.2059$ & 4 & 0.1784 \nl
 0\tablenotemark{A} & $-0.2059$ & 7\tablenotemark{A} & 0.1784 \nl
\enddata
\tablenotetext{A}{The color coefficients for CCDs 0 and 7 are adopted
(median of the other three CCD coefficients).  }
\end{deluxetable}

\clearpage
\begin{figure}
\plotone{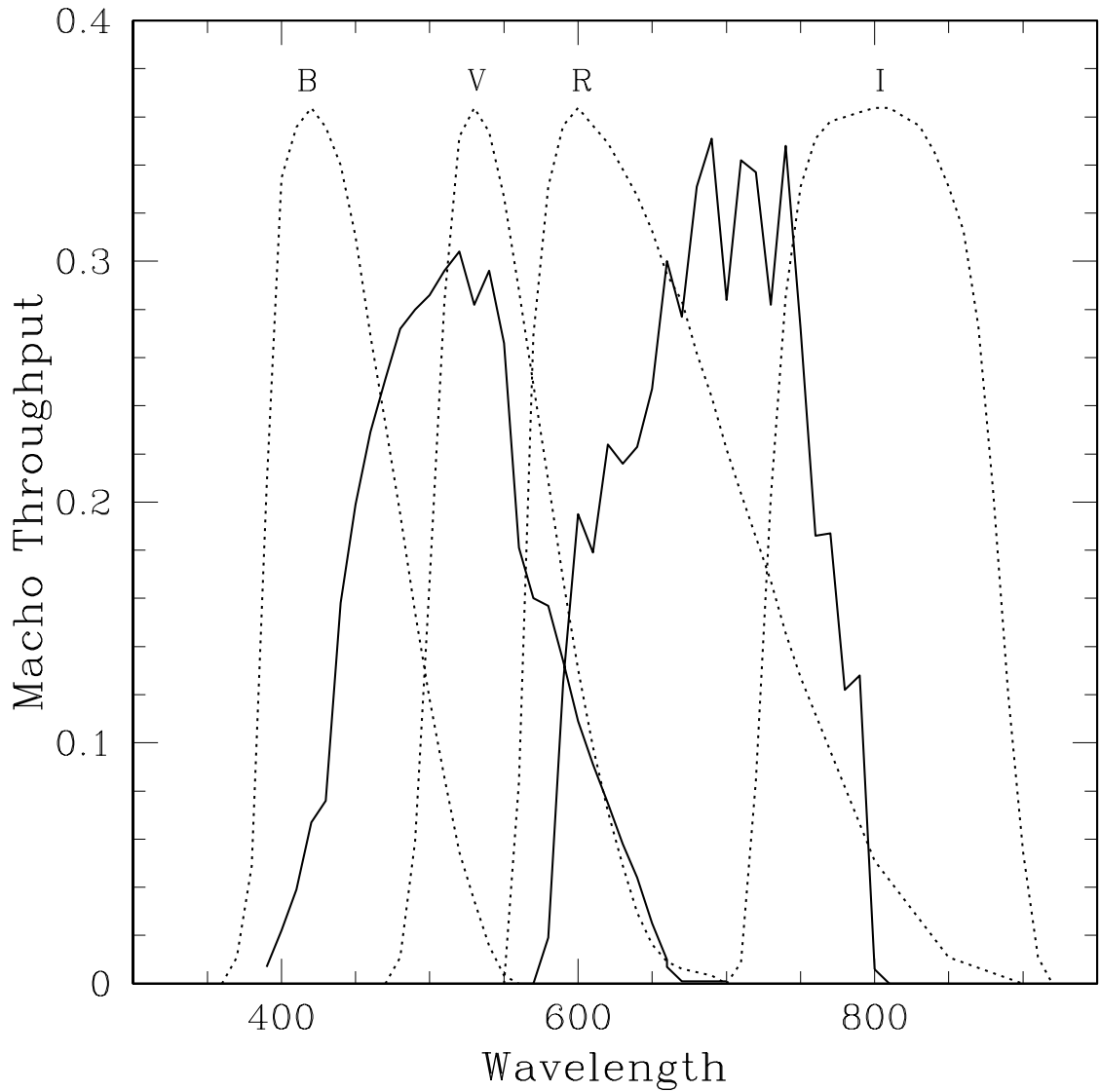}
\caption{Approximate instrumental throughput for the blue and red
MACHO image data.  A throughput of one would indicate no loss of light;
wavelength is in units of $nm$.  
These response functions include the dichroic, filters, and
CCDs. However, the wide-field optics corrector has not been included.  
Uncertainty in these functions is $\sim$20\%.  Also shown are normalized
standard passbands $B$, $V$, $R$ and $I$.}
\end{figure}

\clearpage
\begin{figure}
\plotone{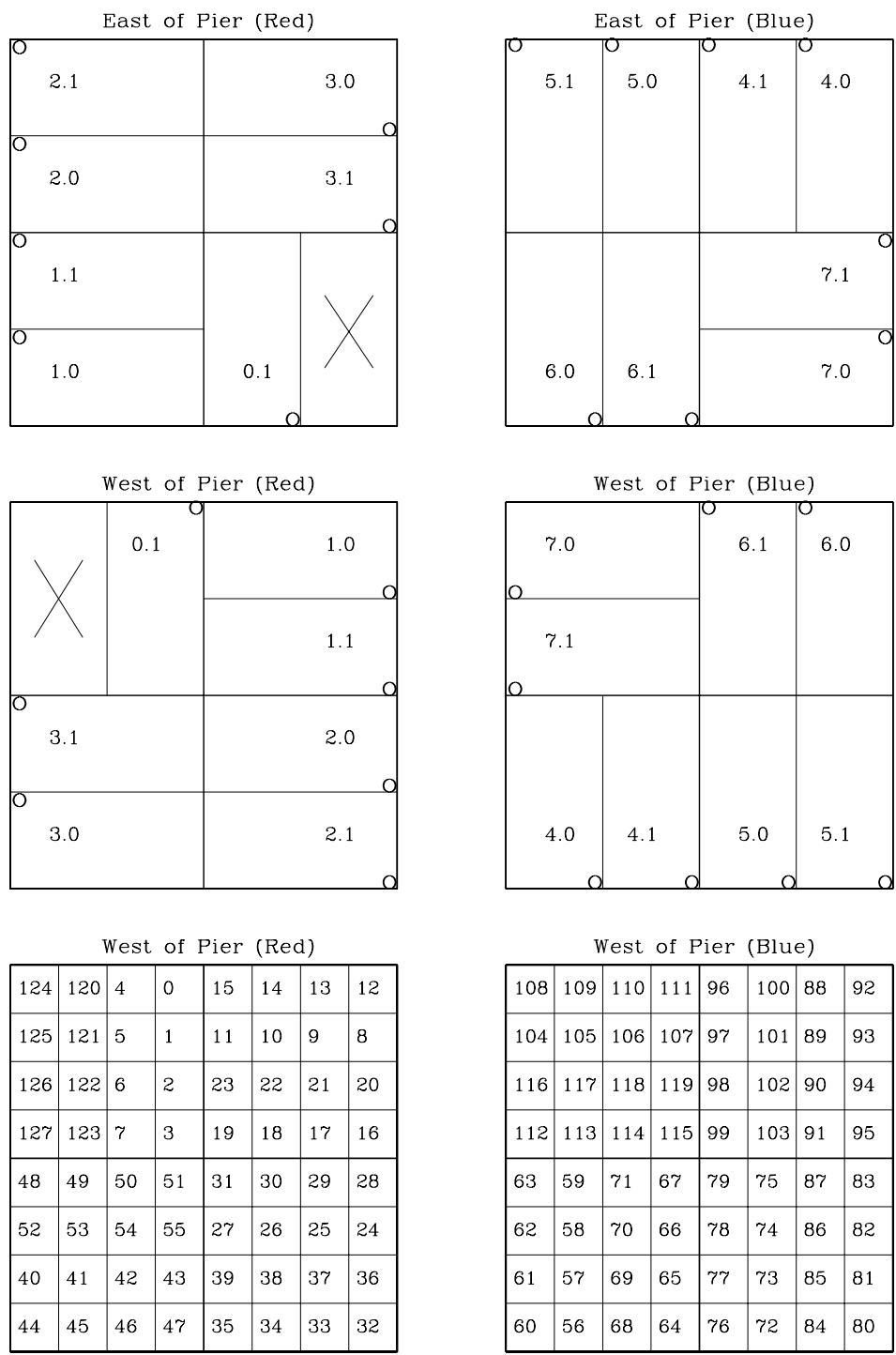}
\caption{Schematic of of the MACHO focal plane (red left, blue right)
in the East of pier (top) and West of pier (middle) orientation.
CCD-Amplifier combinations are labeled.
The inoperational amplifier 0.0 is marked with an X.  
The bottom two panels are the chunk maps of the red (left) and blue (right)
MACHO focal planes in the West of pier orientation.  Chunks are always
fixed to the same CCD-Amplifier.  Thus, the chunk map rotates 180$^{\circ}$ 
when in the East of pier orientation.}
\end{figure}

\clearpage
\begin{figure}
\plotone{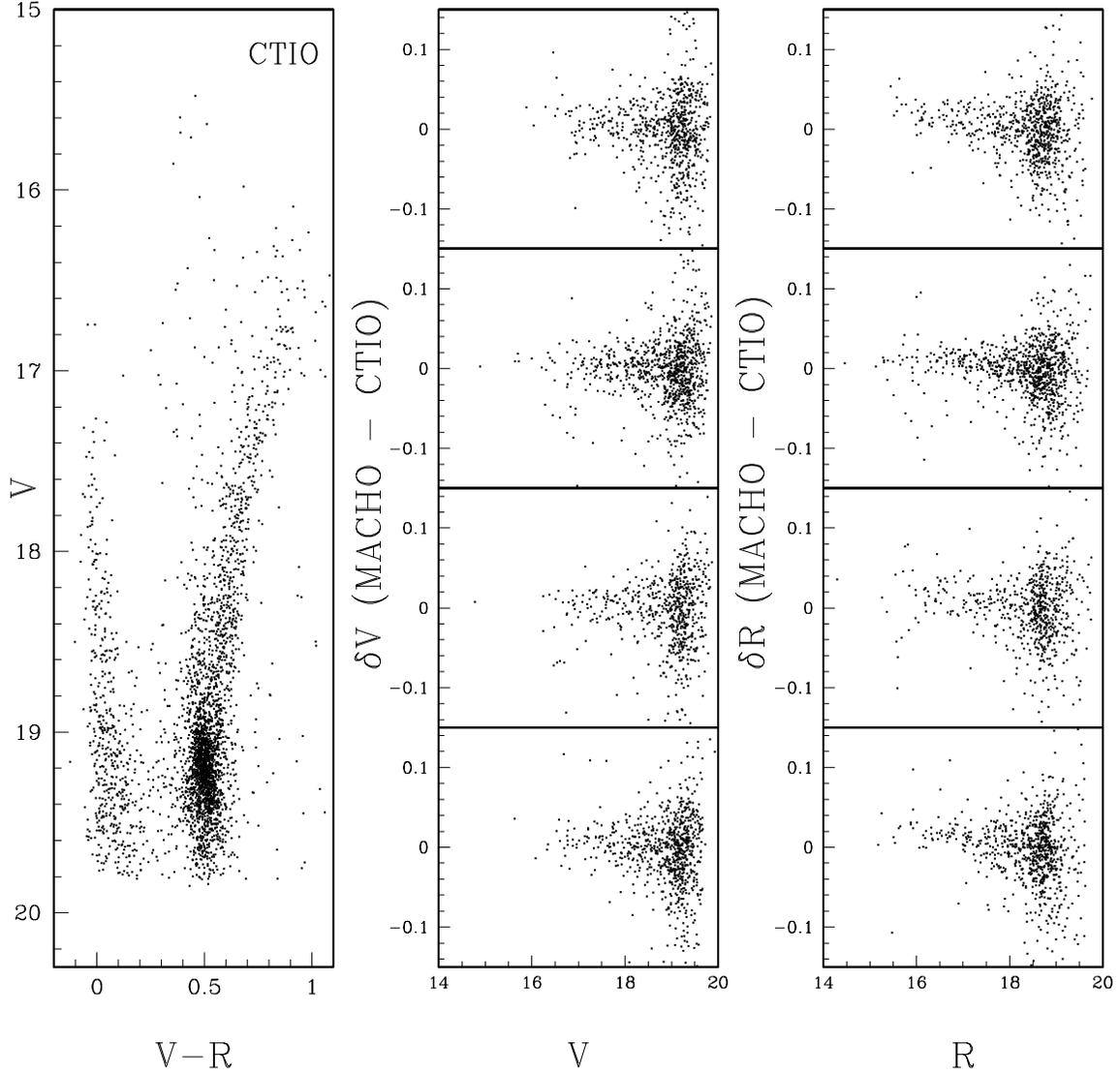}
\caption{CTIO and MACHO calibration data for the four zero-point
chunks in LMC field 13.  The left panel shows the CTIO calibration
data in the CMD; approximately 2500
stars are plotted.  The difference in calibrated
MACHO and CTIO photometry ($\delta V$ and $\delta R$) as a function of $V$
and $R$ mag are plotted in the middle and right panels, respectively.}
\end{figure}

\clearpage
\begin{figure}
\plotone{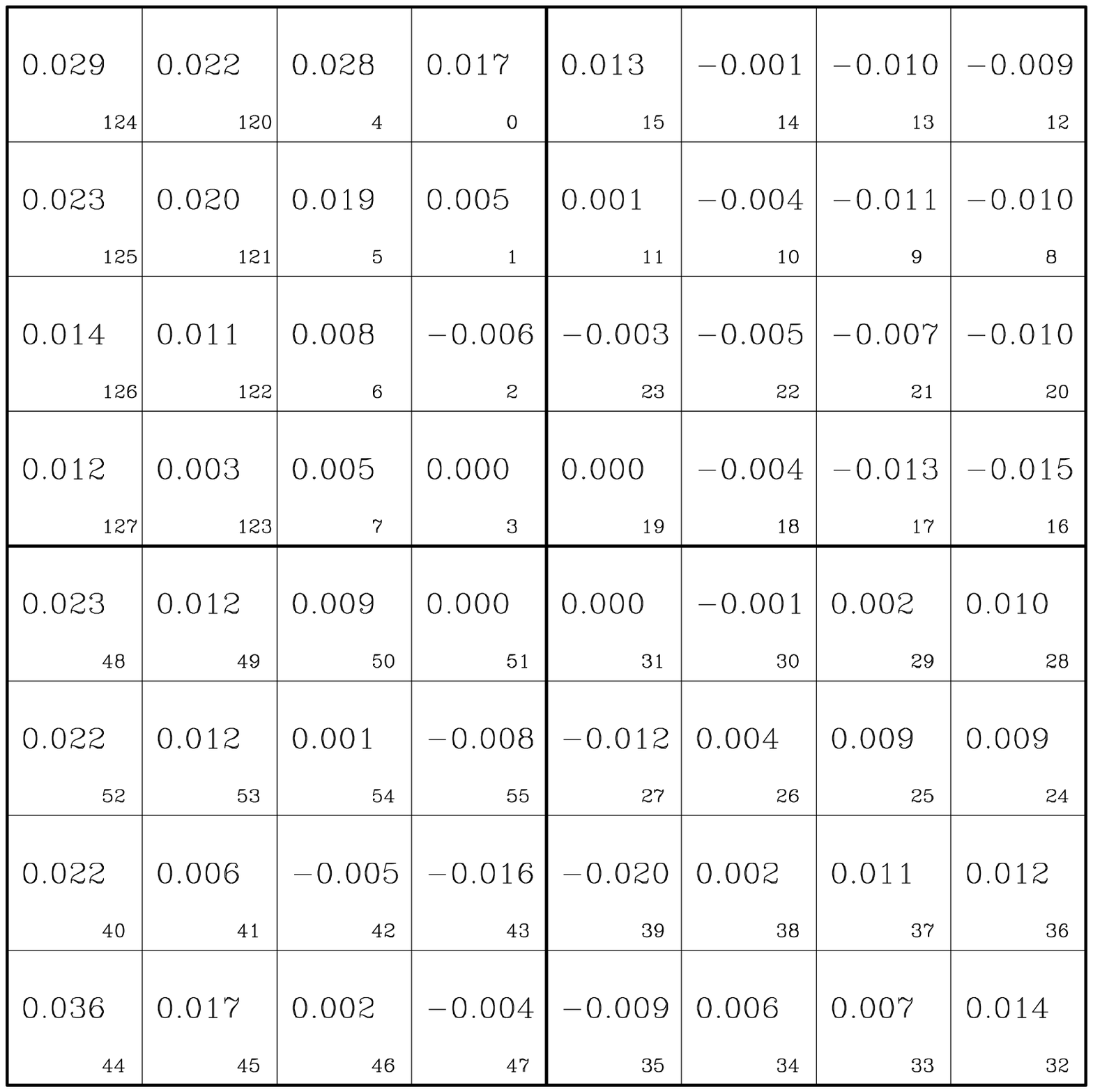}
\caption{Mean chunk offsets calculated for the top-22 field LMC calibration
labeled on a chunk map to illustrate the focal plane dependence. Listed in
each of the 64 chunk locations is mean chunk offset (upper left number),
and red West of pier chunk number (lower right number).  See also Figure~2.}
\end{figure}

\clearpage
\begin{figure}
\plotone{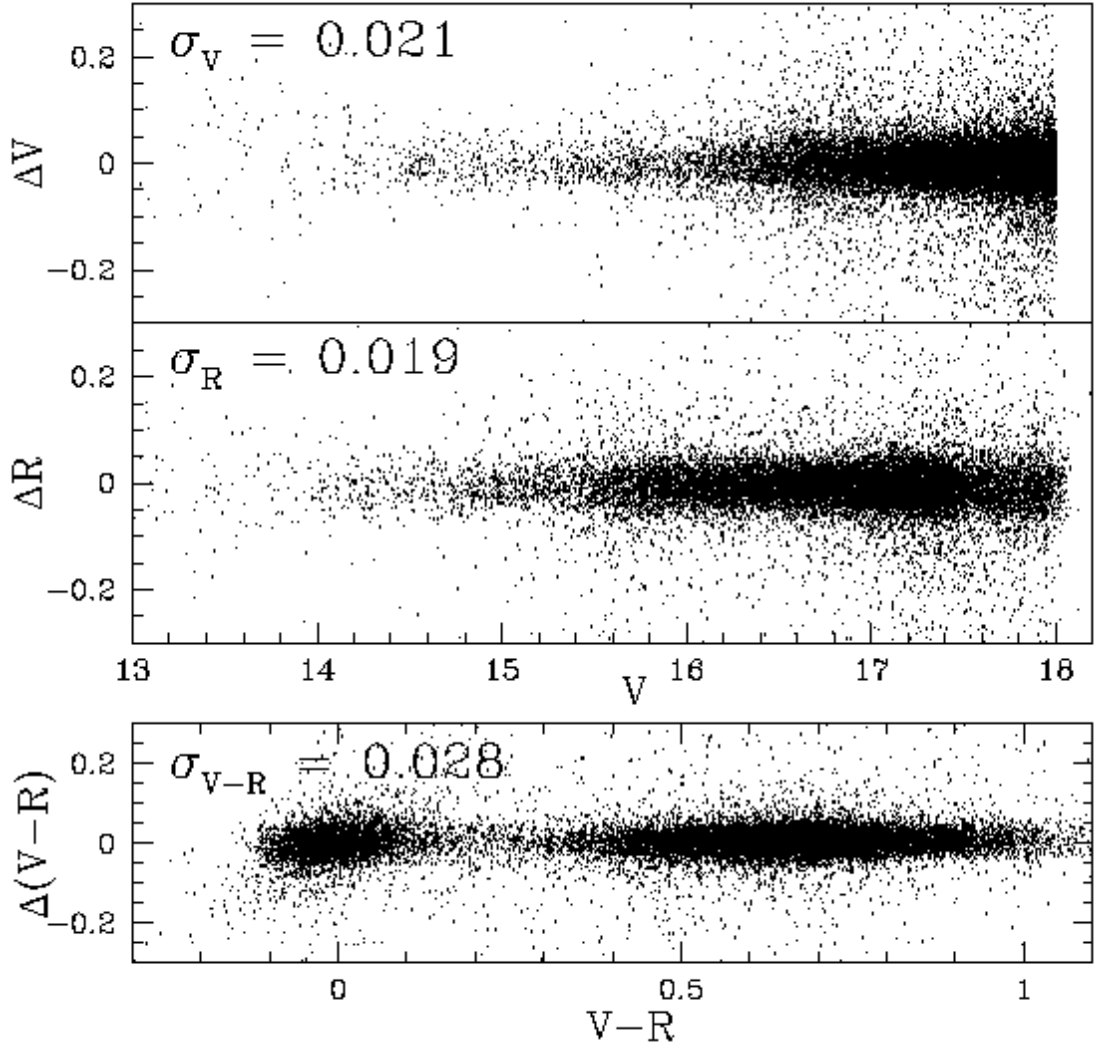}
\caption{The differences in $V$, $R$, and $(V-R)$ for stars in field-overlap
regions are plotted as a function of magnitude or color in the
top, middle, and bottom panels respectively.  Approximately 20,000
stars are plotted, each have $V \simlt$ 18 mag and are located
in 150 chunks tying together 21 of the LMC top-22 fields.
The pair-wise comparisons of overlapping chunk photometry are
plotted in no particular order in a global sense, but all chunk
pairs connecting the same two fields have been subtracted in the
same order.
Standard deviations,
as described in the text and illustrated
here, are labeled in each panel.  These values indicate the precision of the
calibrated MACHO photometry for $\sim$9 million stars distributed over
10 square degrees of the LMC bar.}
\end{figure}

\clearpage
\begin{figure}
\plotone{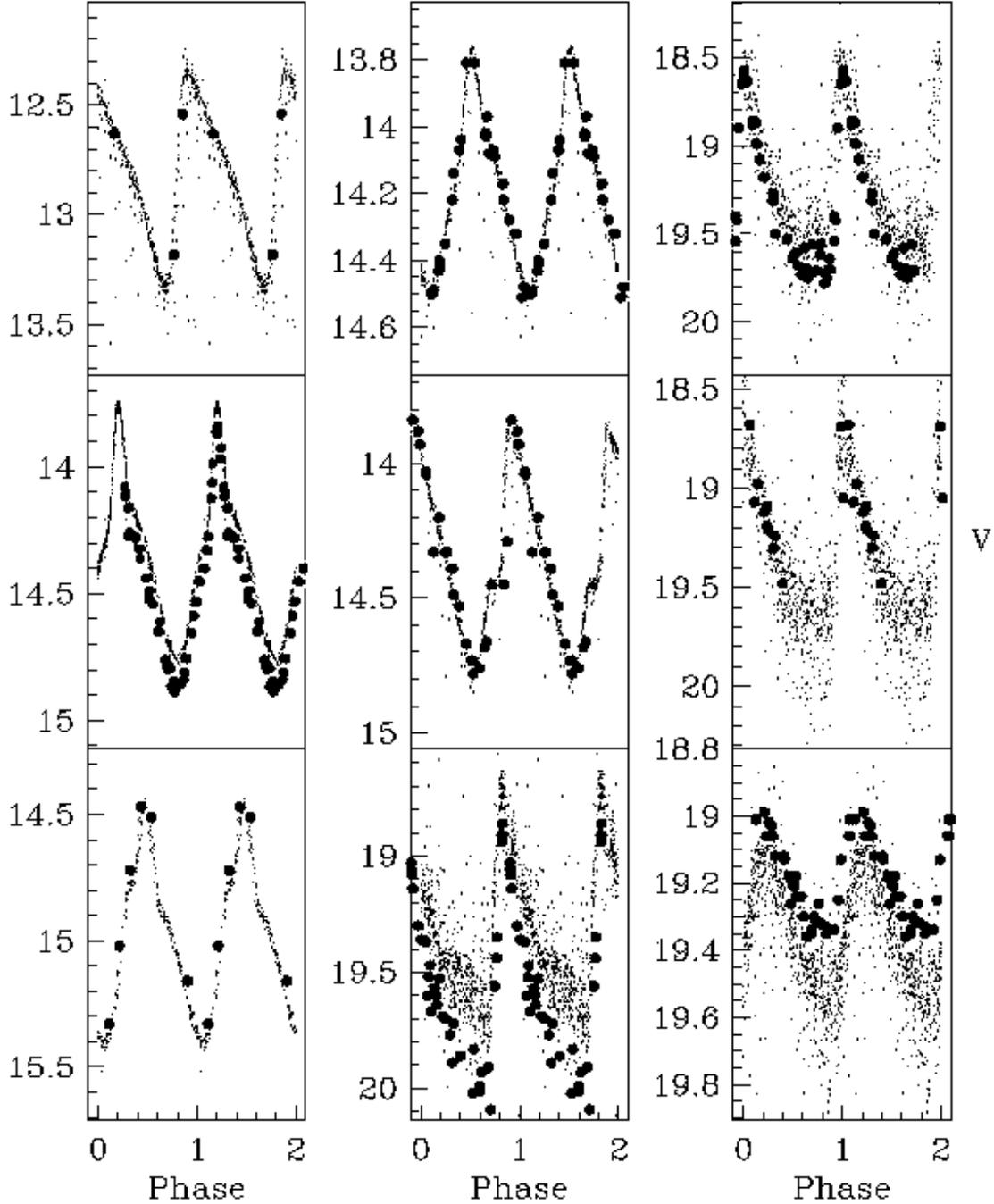}
\caption{Period-folded MACHO $V$ lightcurves compared with 
photometry data assembled from the literature (5 different sources). 
MACHO data are plotted as dots and the comparison data as
filled circles.  Error bars are omitted for clarity.  From top to bottom
then left to right the variables are Cepheids: 
HV900, HV905, HV2510, HV2352, HV2324, and RR Lyrae near the cluster NGC~1835:
GR-6 GR-14, GR-16, and Walker-V26.  See text for further details.}
\end{figure}

\clearpage
\begin{figure}
\plotone{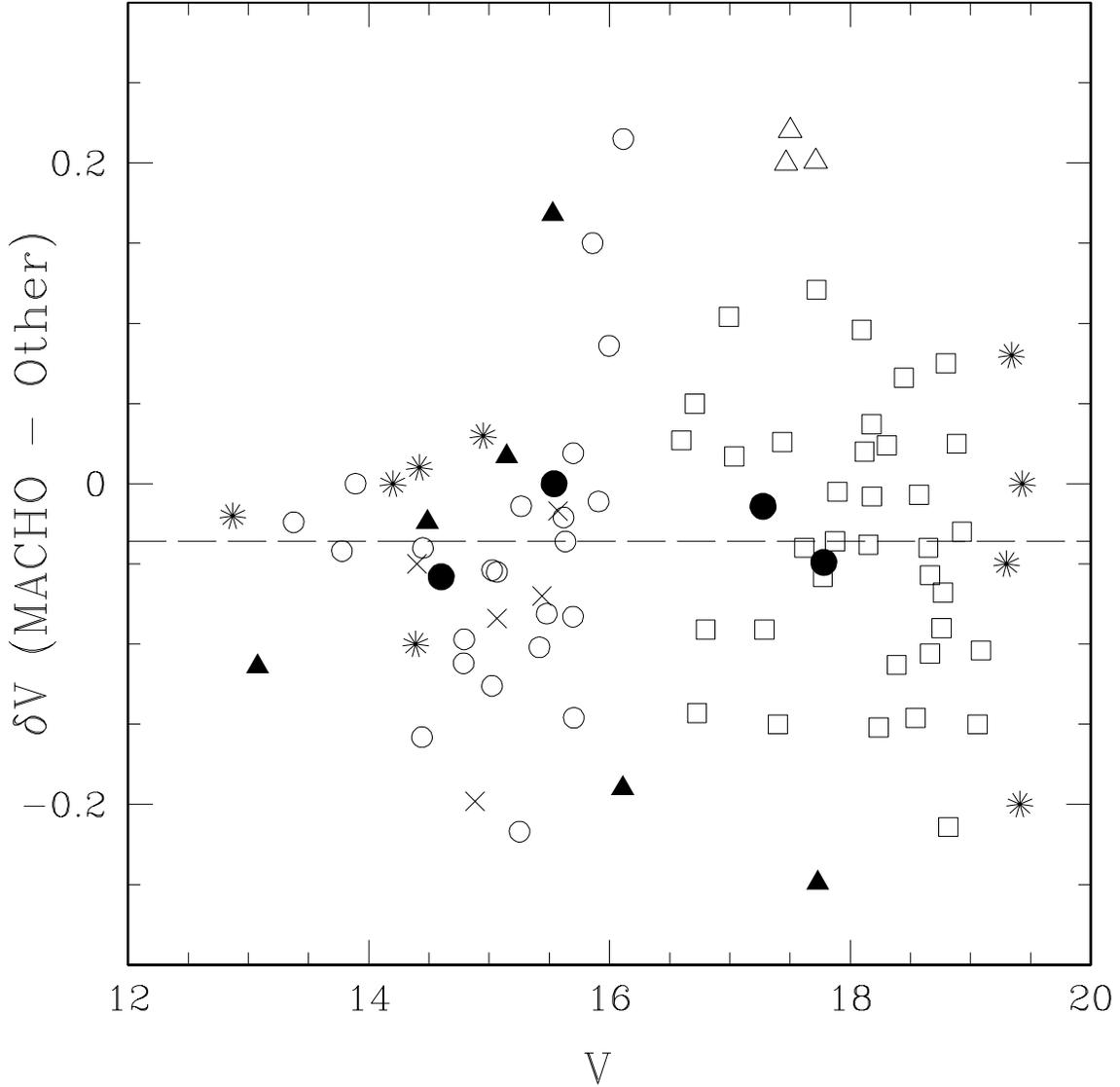}
\caption{MACHO calibrated $V$ photometry in the 
LMC compared with various published measurements.
The data are represented by the following symbols: 
asterisks are the Cepheids
and RR Lyrae from Fig.~6, 
filled circles are Walker's standard star sequence near
NGC~1835 (1993), 
open triangles are standard sequence of Cowley et al.~(1990) near Cal-87,
filled triangles are data from Flower et al. (1982) near NGC~2058/2065,
open squares are stars
near NGC~1847 from Nelson and Hodge (1983), and the open circles are
photometry assembled
from the classical LMC photometry paper by Tifft and Snell (1971).
The median offset between the MACHO and all of the other data is
indicated with a dashed line.}
\end{figure}

\clearpage
\begin{figure}
\plotone{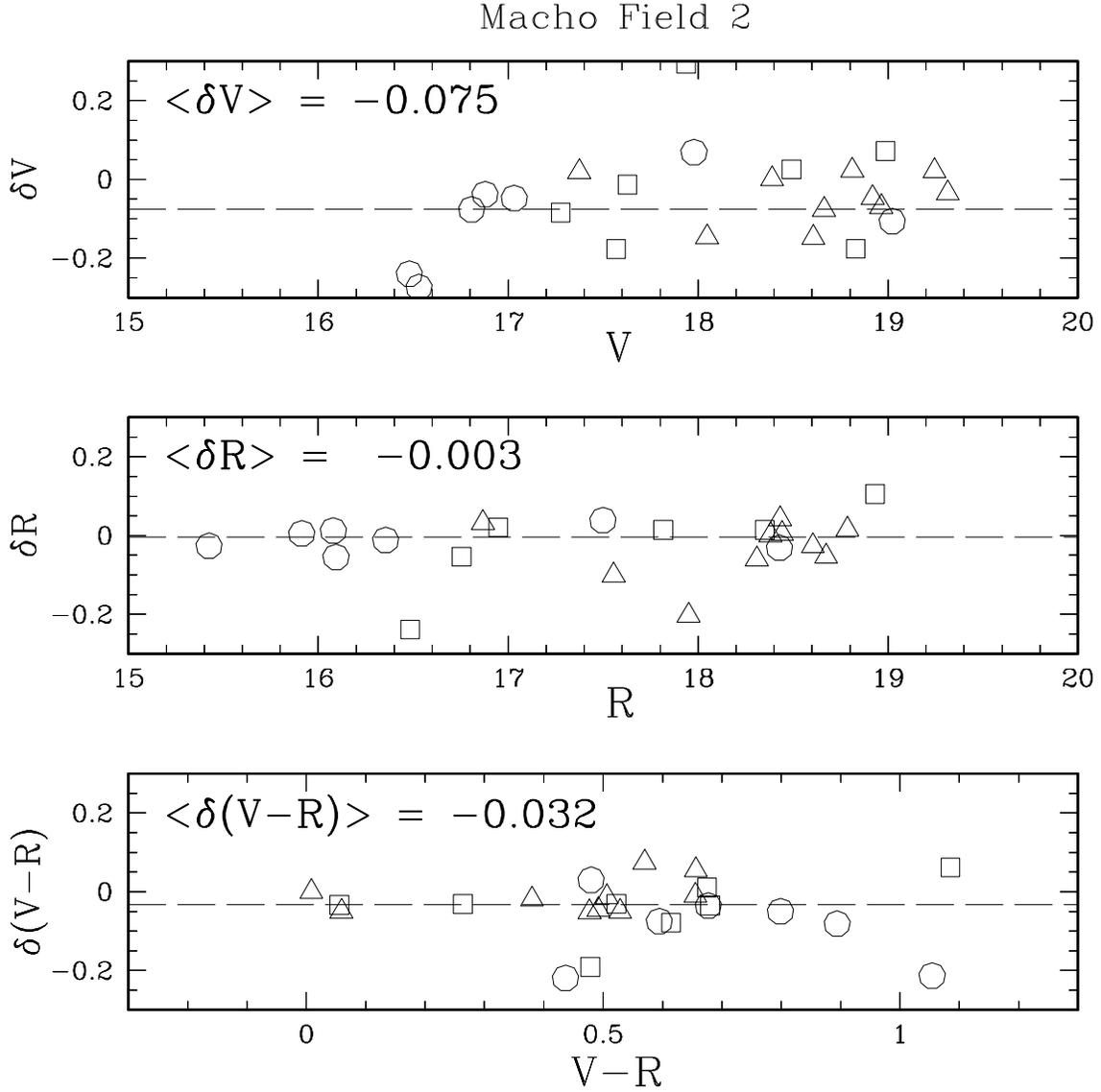}
\caption{MACHO calibrated photometry for field 2 compared with
HST calibrated photometry.  Open trangles are data from the WF2 chip,
open circles from WF3, and open squares from WF4. 
We plot the magnitude or color offset, $\delta V$, $\delta R$,
and $\delta(V-R)$, in the sense (MACHO - HST) versus magnitude
or color.  The median offset is labeled in each panel.}
\end{figure}

\clearpage
\begin{figure}
\plotone{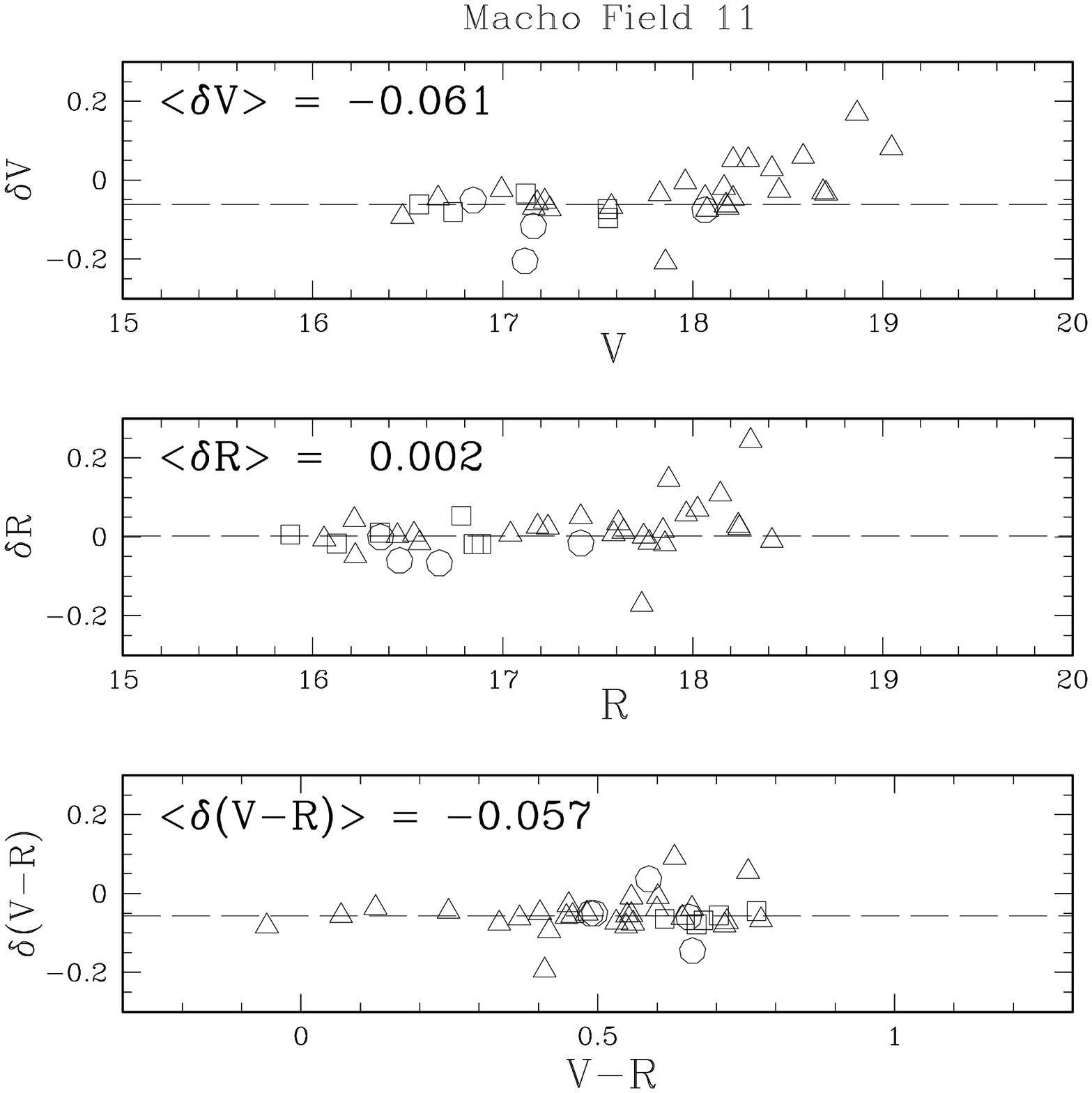}
\caption{Same as Fig.~8, but for MACHO field 11.}
\end{figure}

\clearpage
\begin{figure}
\plotone{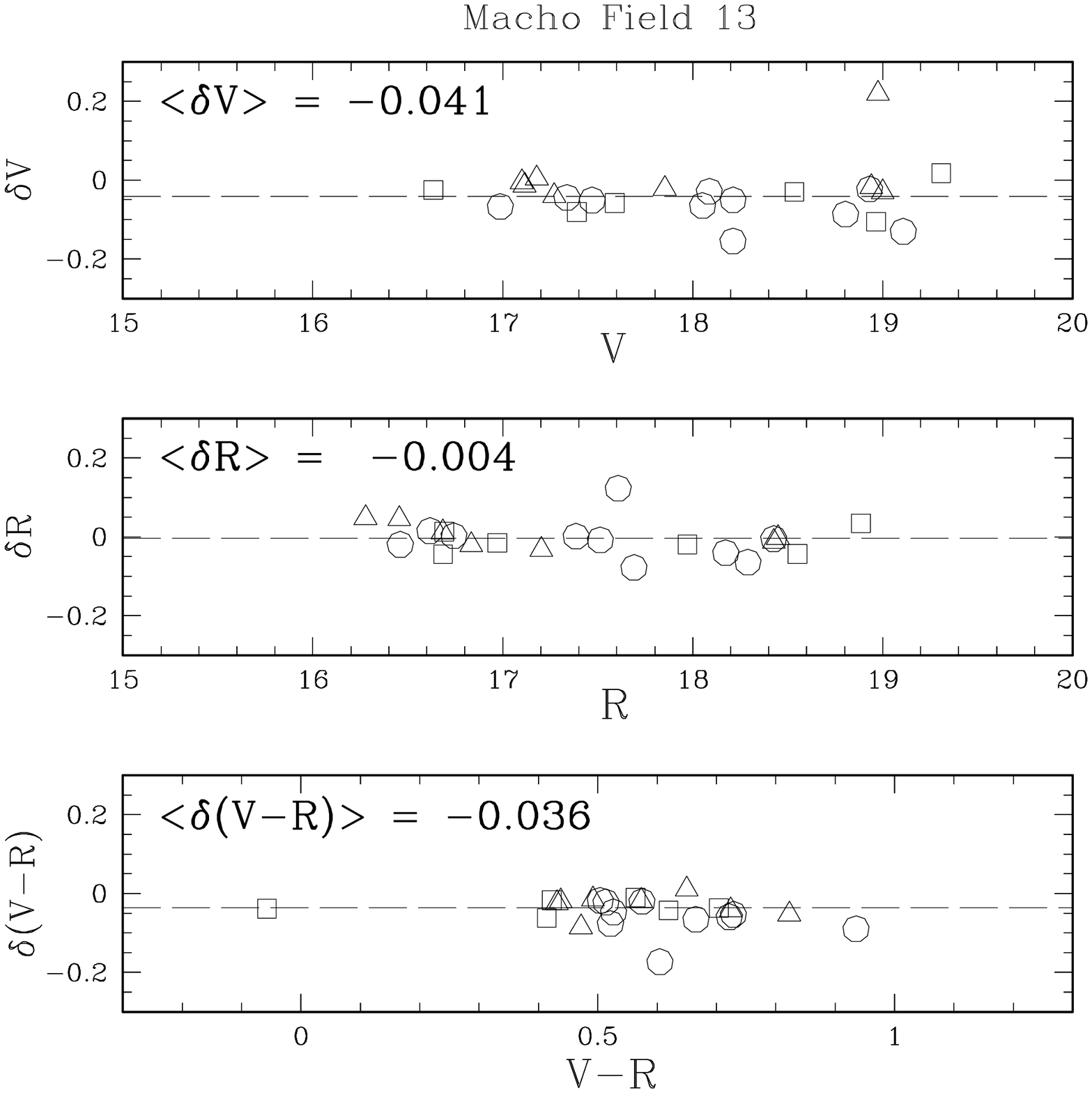}
\caption{Same as Fig.~8, but for MACHO field 13.}
\end{figure}

\clearpage
\begin{figure}
\plotone{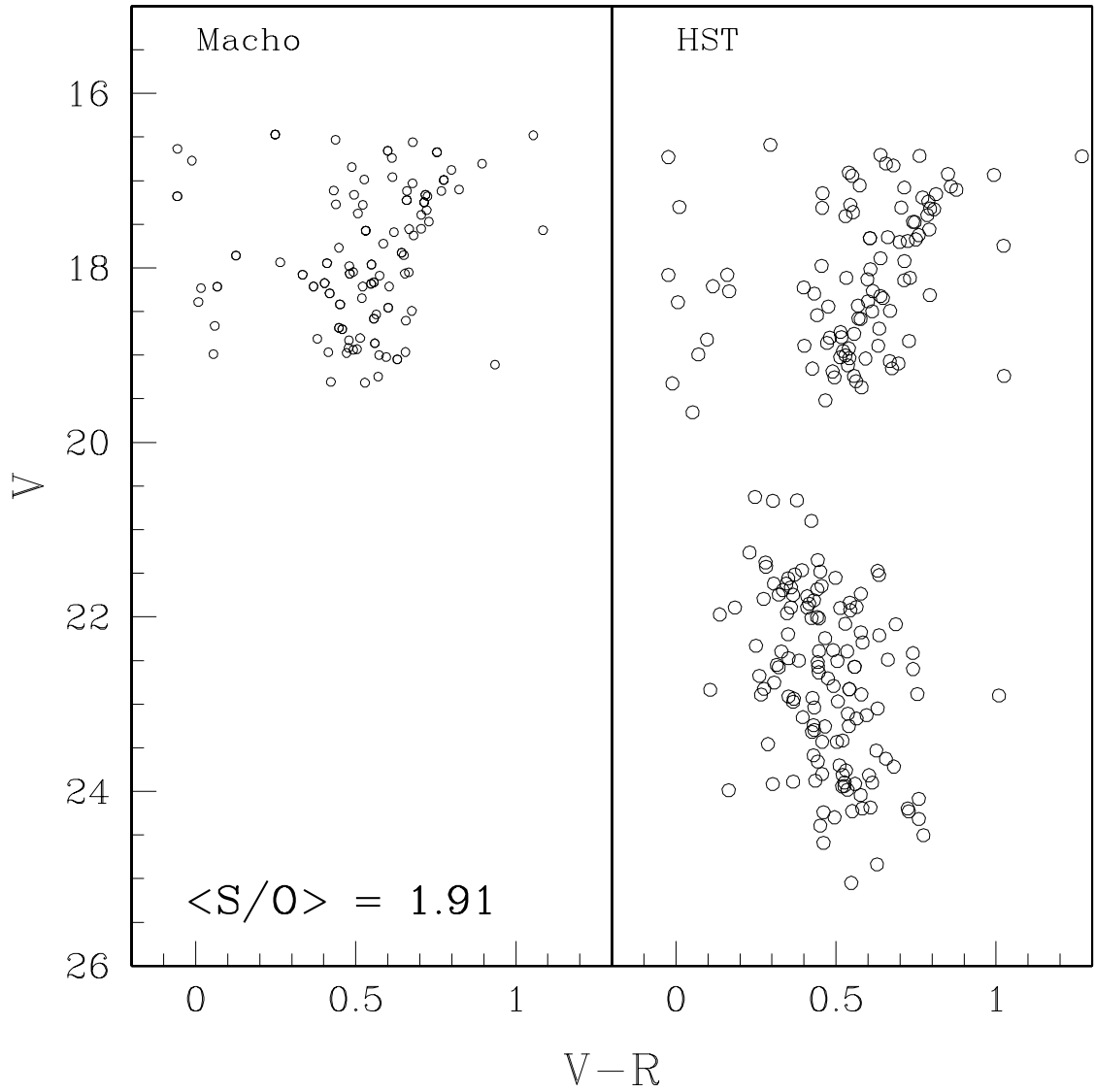}
\caption{Color-magnitude diagrams showing the MACHO objects and
(un-blended) HST stars from Figures 8-10.  Each panel is labeled.
The ratio of stars to objects is 229/120 = 1.91, also labeled.} 
\end{figure}

\clearpage
\begin{figure}
\plotone{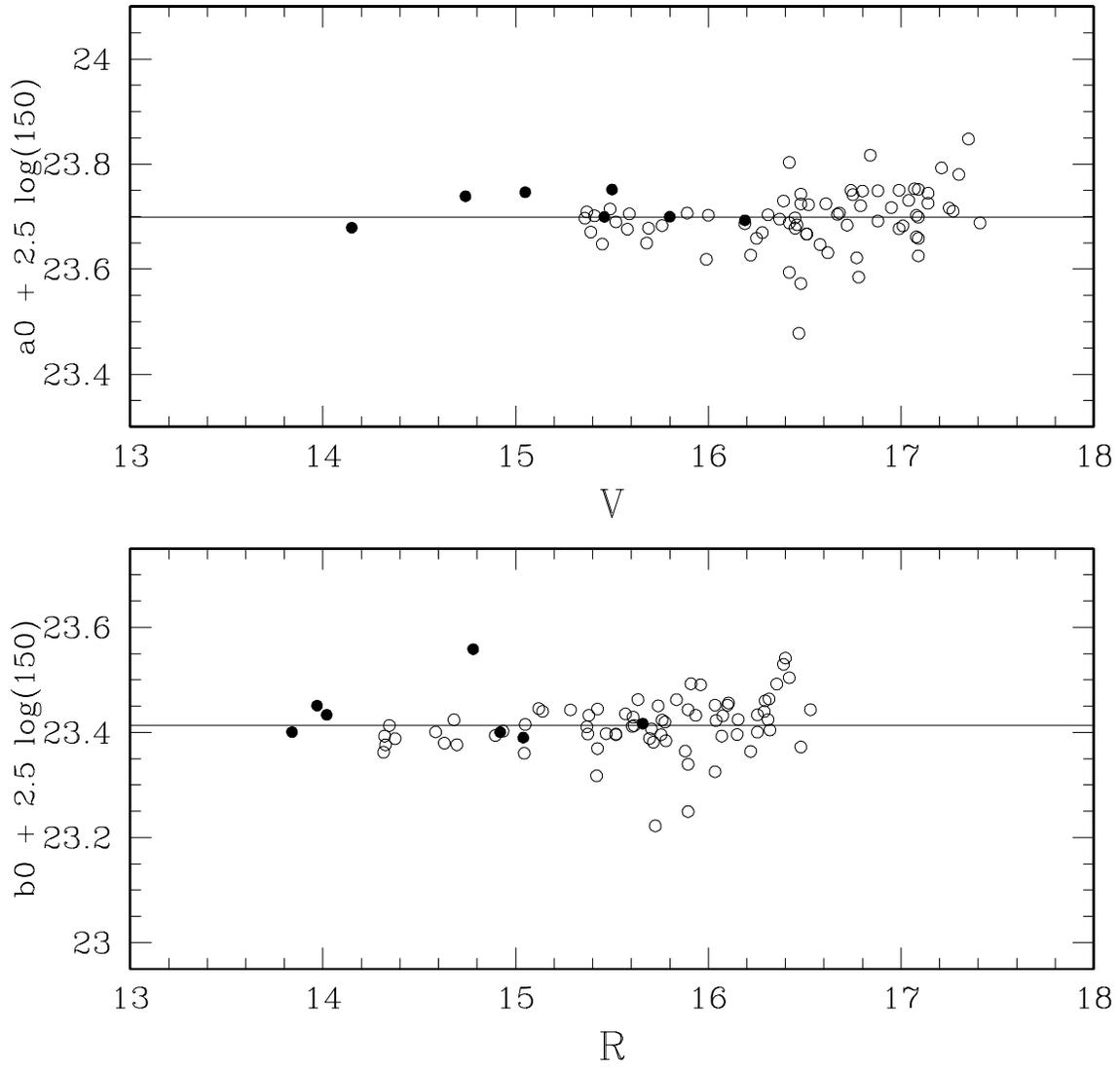}
\caption{The solution for $a0$ and $b0$ using the
published photometry of Cook (1986) and Walker and Mack (1990), plotted as
open and filled circles respectively, in 
MACHO bulge field 119.}
\end{figure}

\clearpage
\begin{figure}
\plotone{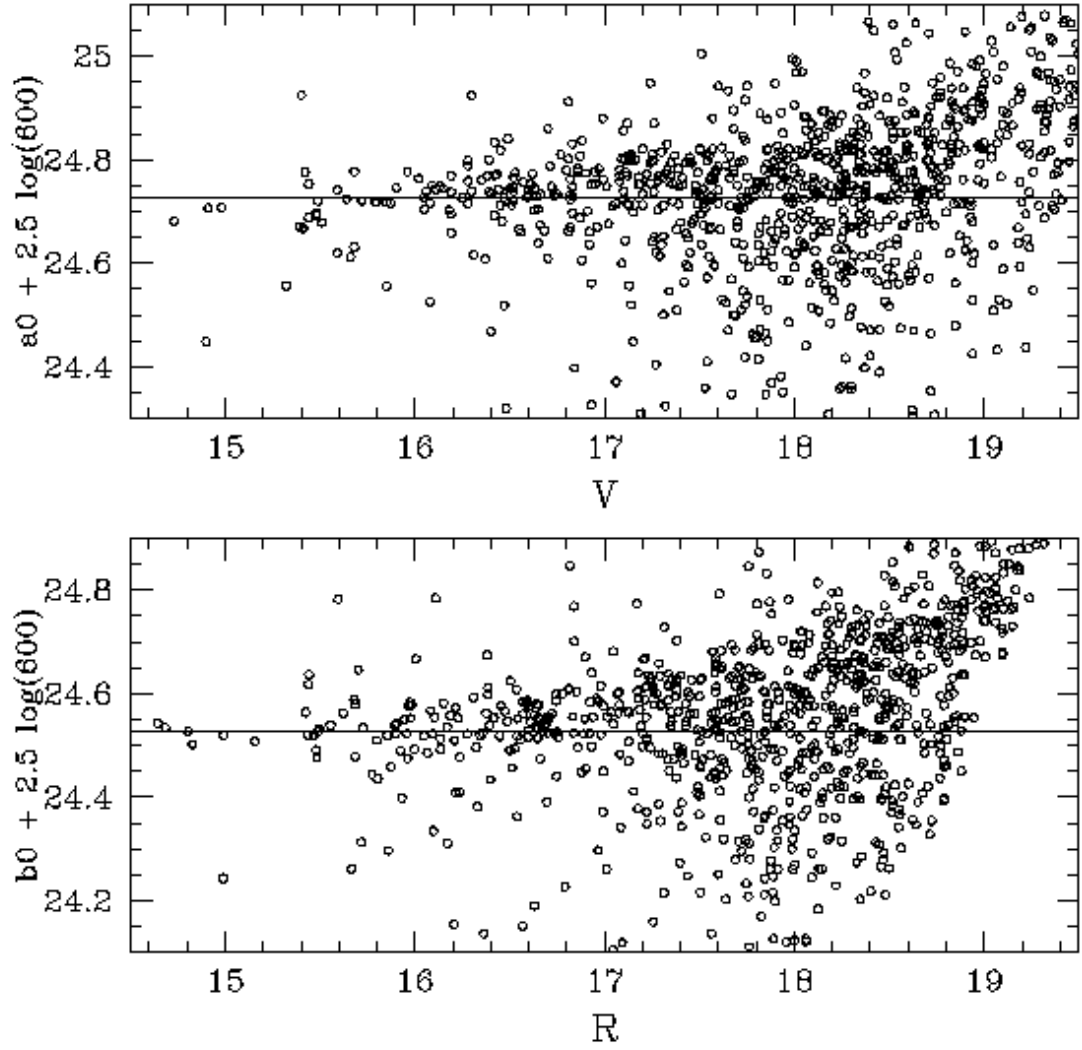}
\caption{The solution for $a0$ and $b0$ using the
published photometry of Vallenari et al.~(1994) in 
MACHO SMC field 207.}
\end{figure}

\clearpage
\begin{figure}
\plotone{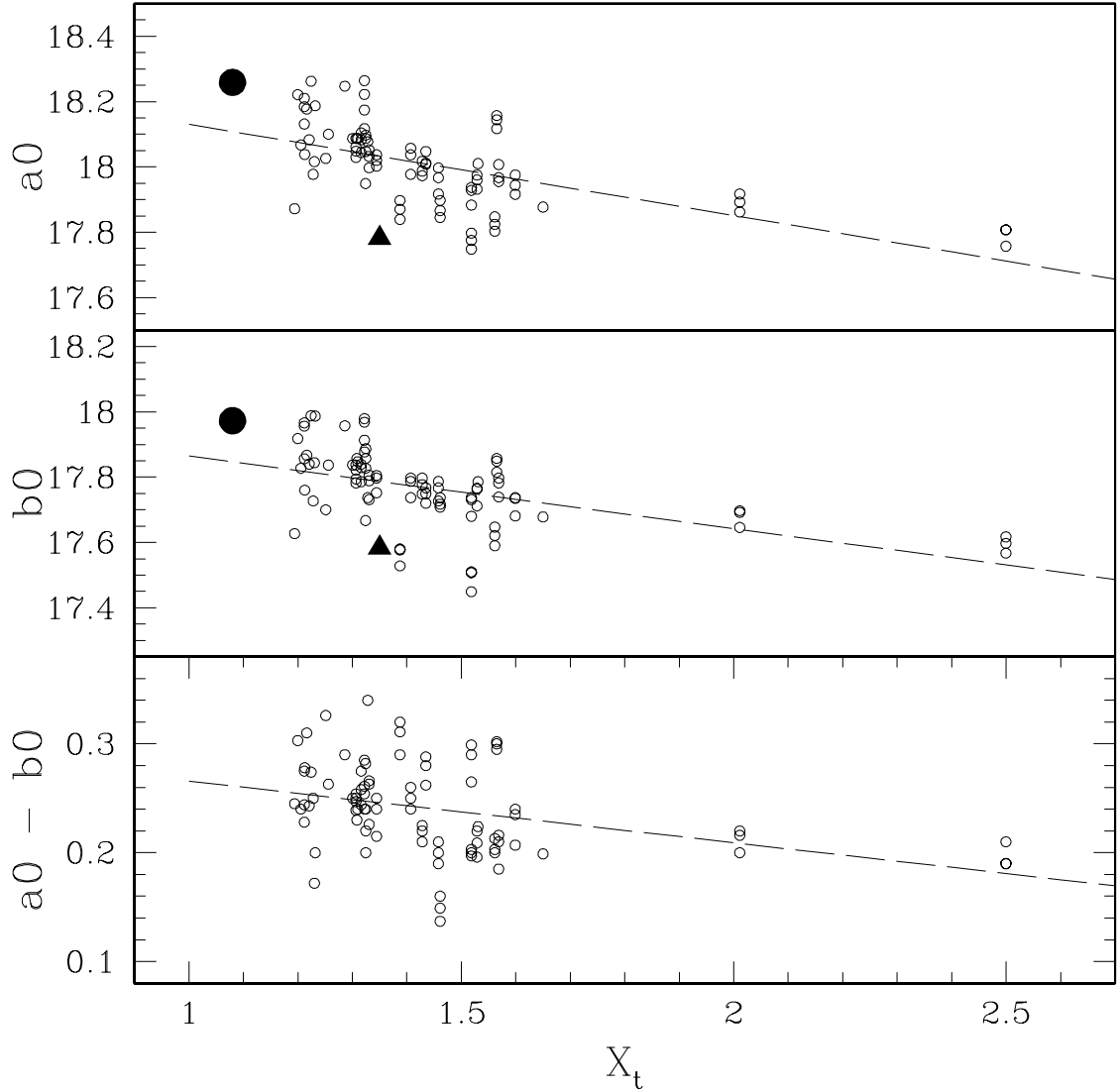}
\caption{Calibration zero-points ($a0$ and $b0$) for the top-22 LMC fields
(open circles),  Galactic bulge field 119 (filled circle), and SMC field
207 (filled triangle) plotted as a function of template airmass, $X_t$.
Dashed lines show the regressions used to predict the calibrations zero-points
for all other MACHO fields.}
\end{figure}

\clearpage
\begin{figure}
\plotone{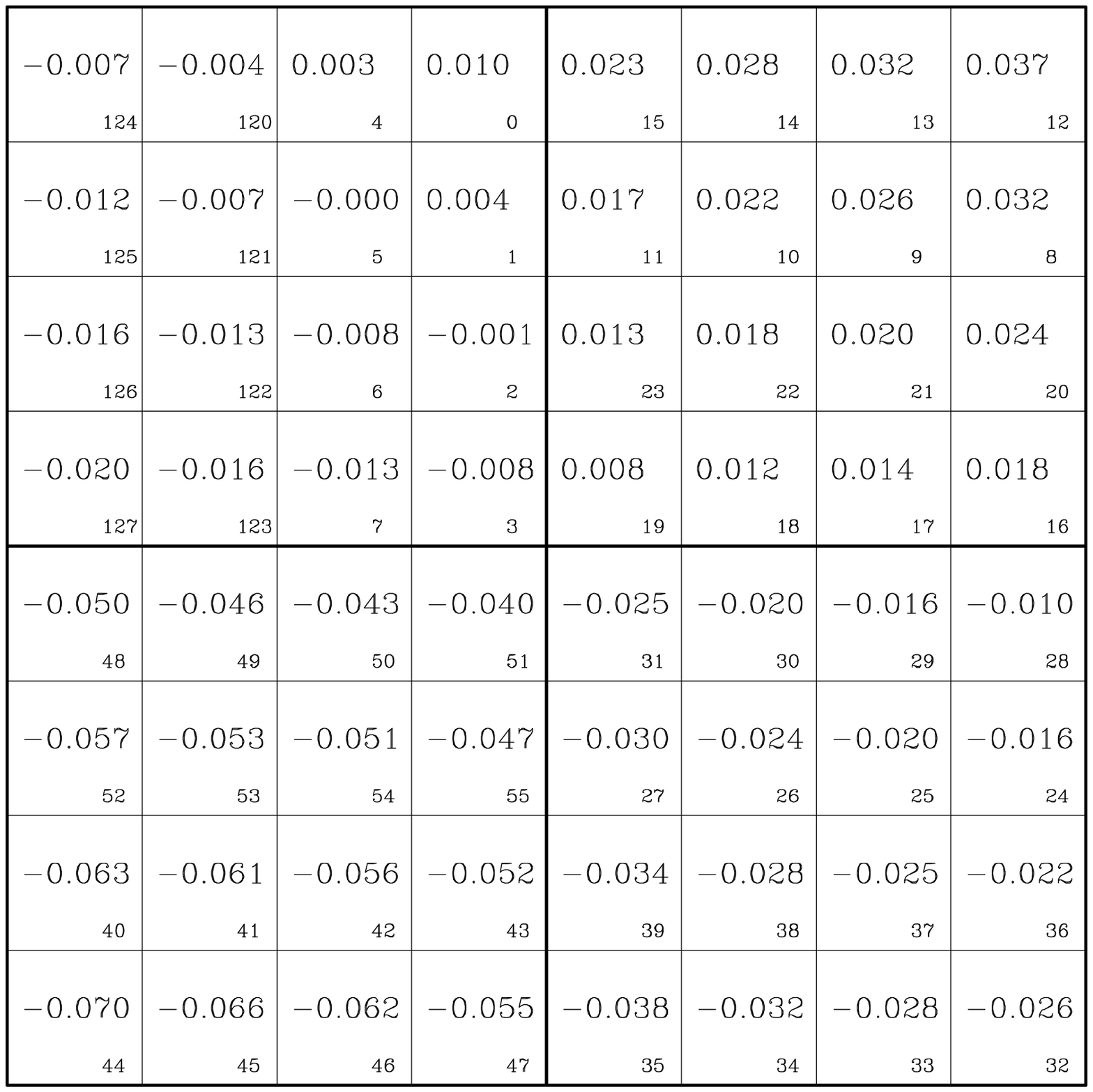}
\caption{Blue jitter $BJ_{w}$ coefficients 
labeled on a chunk map to illustrate the focal plane dependence.  Center
in each of the 64 chunk locations is $BJ_{w}$ while the red West of pier
chunk number is the lower right number (see also Figure~2).  These coefficients
are for blue jitter corrections in the imaginary case of a 
West of pier observation made of a field with template photometry derived from
an entirely East of pier observation.}
\end{figure}

\clearpage
\begin{figure}
\plotone{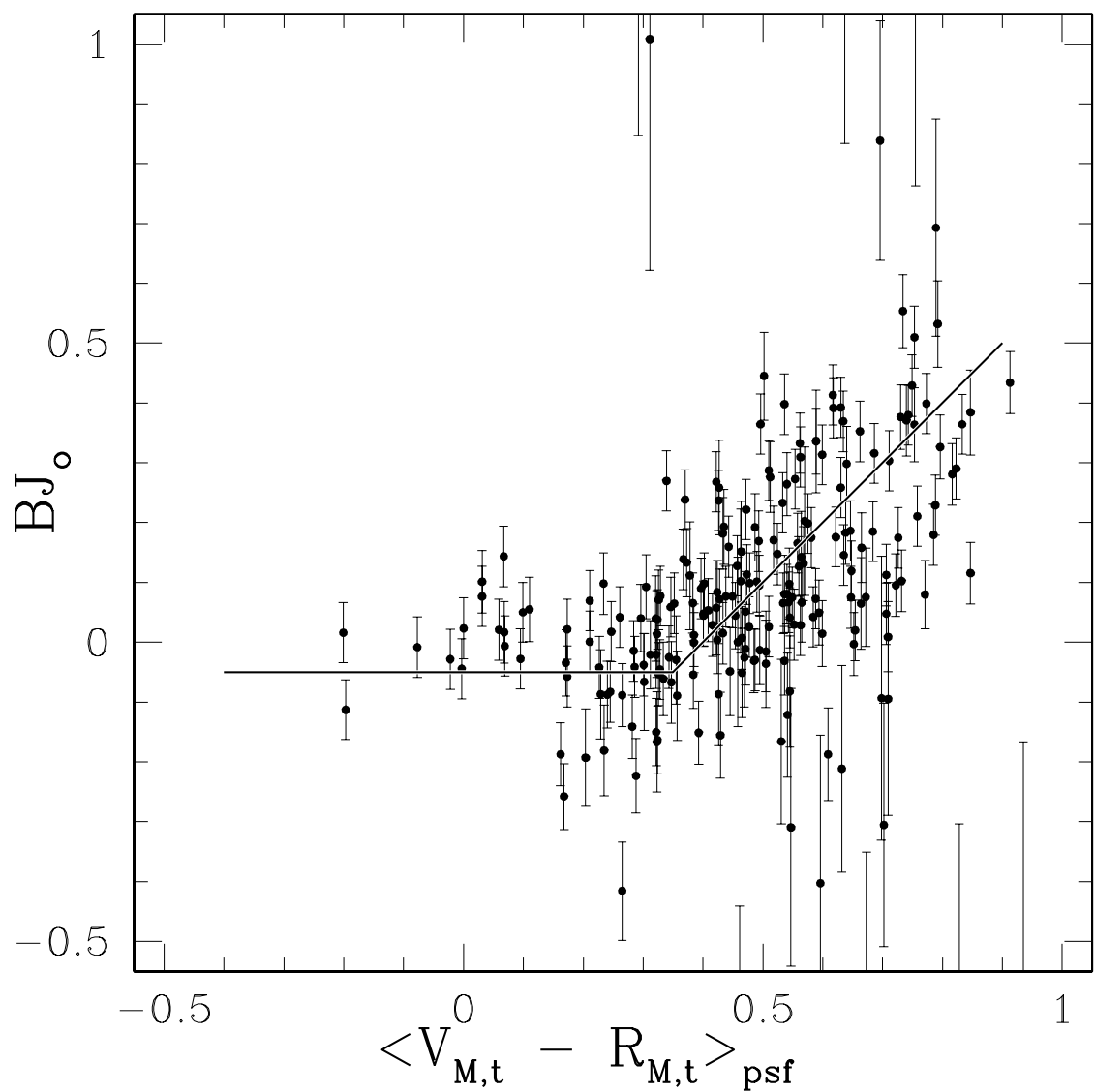}
\caption{Data for 224 chunks in 5 different LMC fields used to
derive the $BJ_o$--(mean color of PSF stars) calibration for the 
blue jitter corrections.  The solid line is the adopted calibration.} 
\end{figure}

\clearpage
\begin{figure}
\plotone{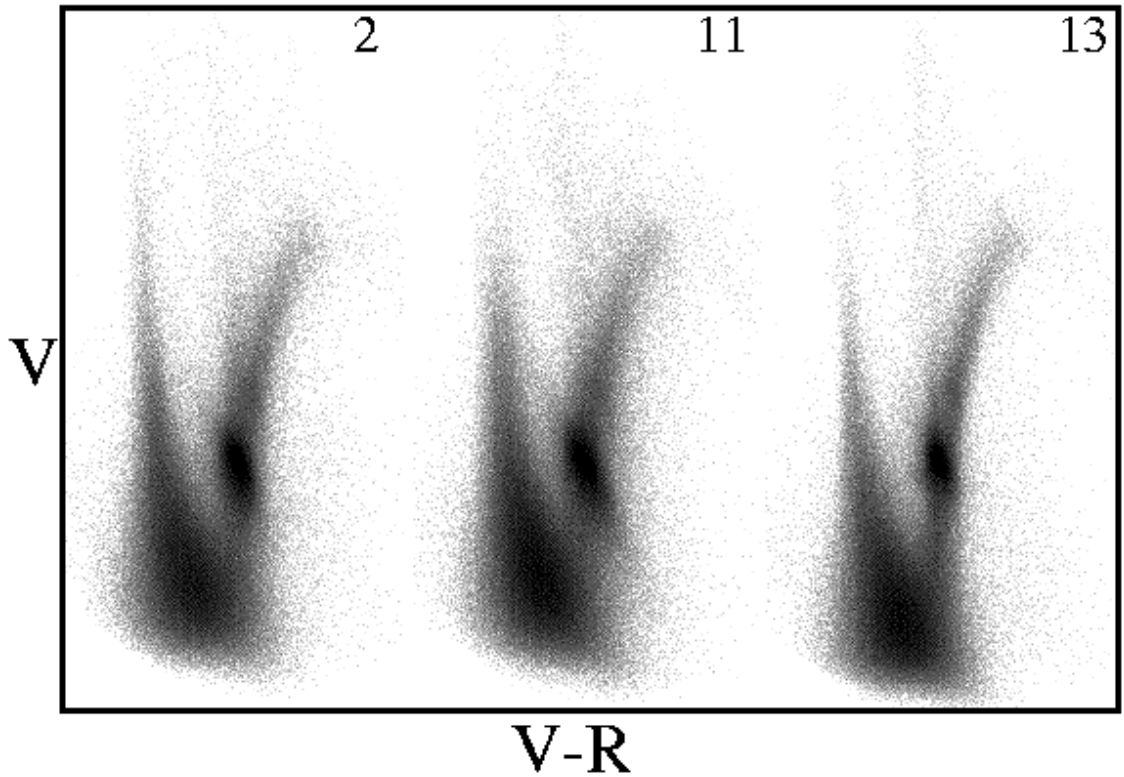}
\caption{MACHO calibrated Hess diagrams for LMC fields 2, 11, and 13.
Although axes are not labeled, for each field they run from 
$(V-R) = -0.5$ to 1.5 mag (bin size 0.01 mag), 
and $V = 22$ to 14 (bin size 0.02).  Intensity represents the number of stars.
The diagrams have been log-scaled.  These diagrams allow a qualitative
comparison of the CMDs for each field.} 
\end{figure}

\clearpage
\begin{figure}
\plotone{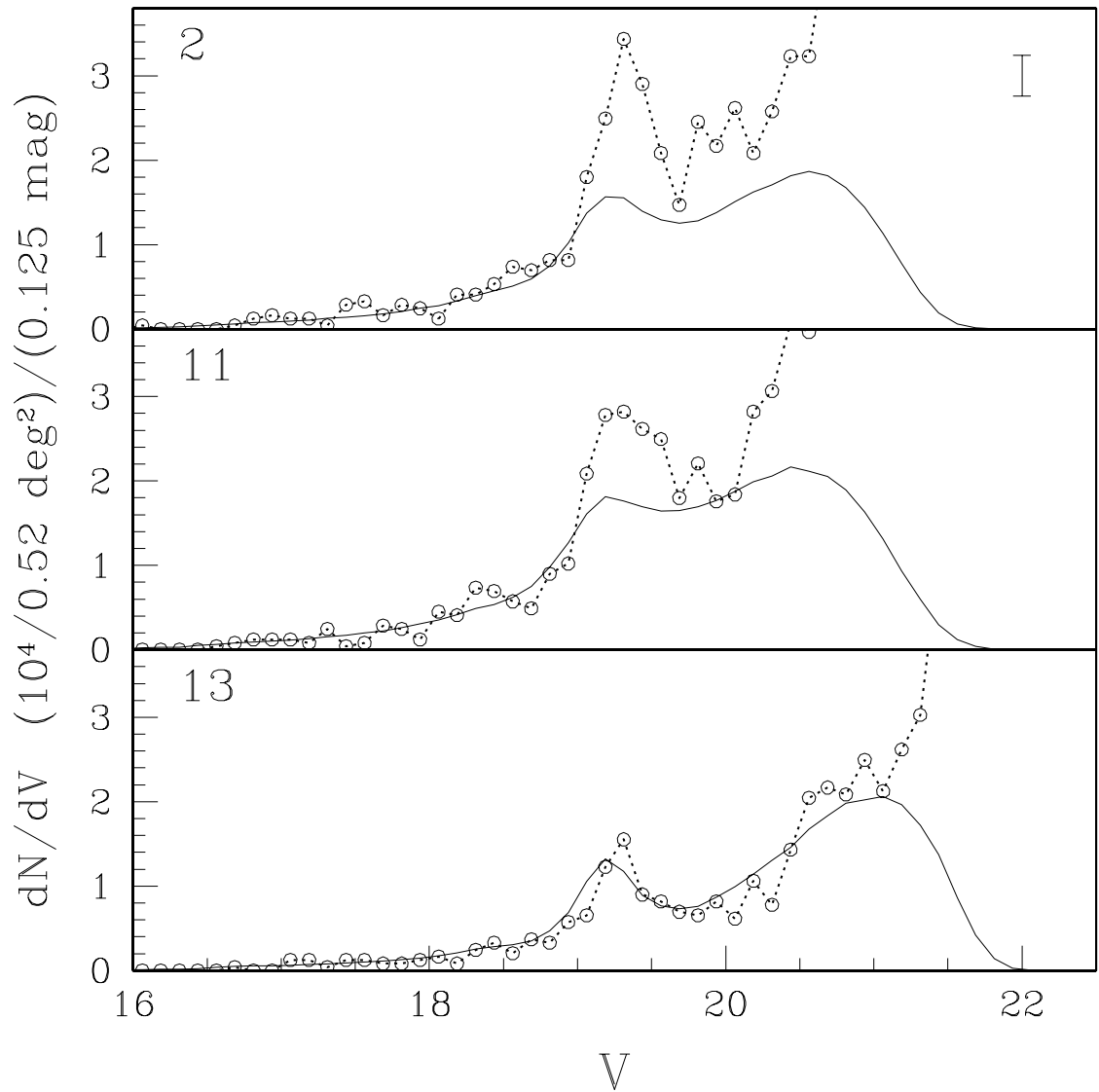}
\caption{Luminosity functions for LMC fields 2, 11, and 13
comparing MACHO and HST data. The HST data has been scaled to the MACHO
data.  The typical errorbar associated with each bin of HST data is
indicated in the upper right corner.}
\end{figure}

\clearpage
\begin{figure}
\plotone{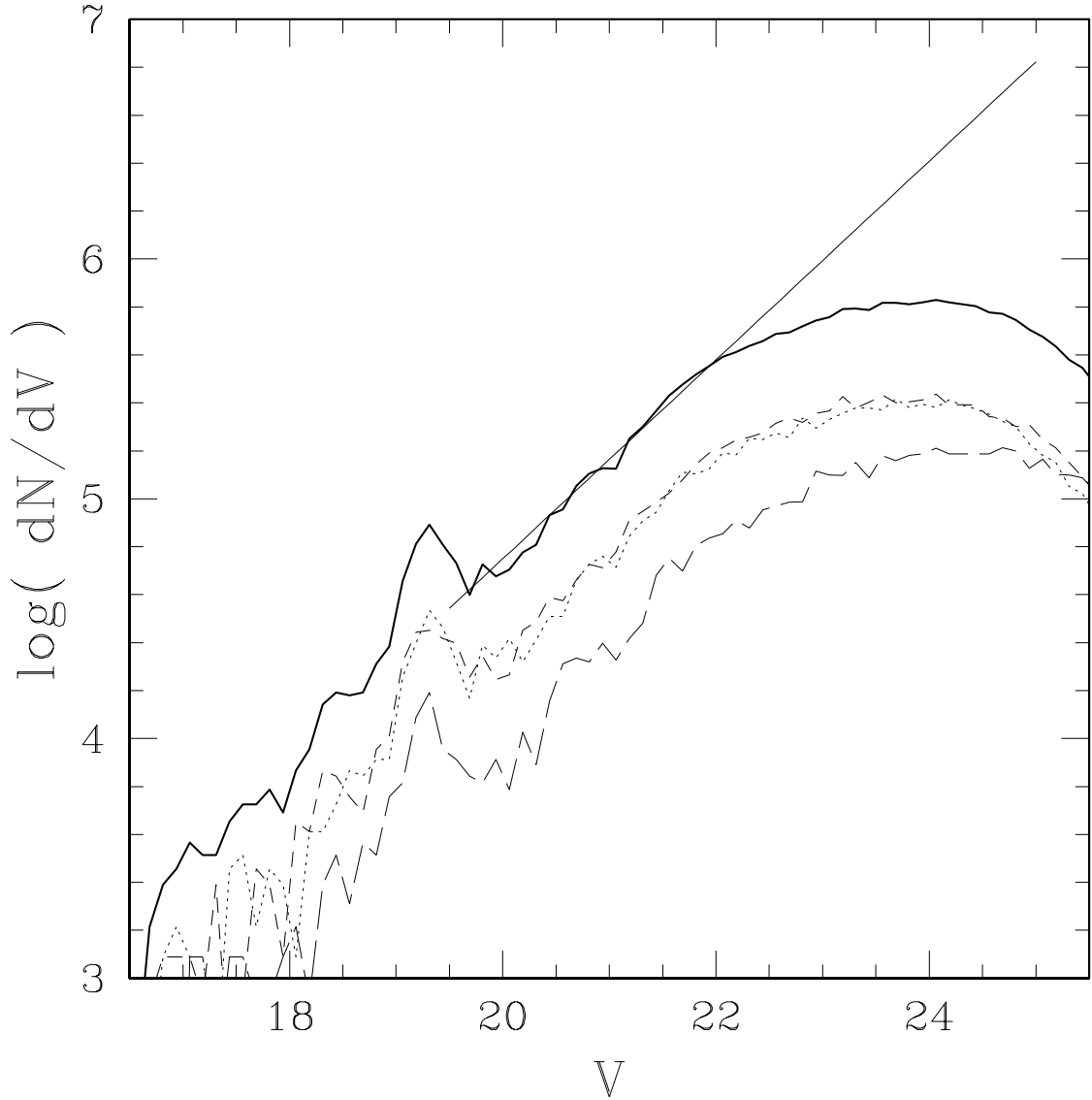}
\caption{Luminosity functions for LMC fields 2, 11, and 13
comparing HST data.  The units of $dN/dV$ are as in Fig.~18.
The sum of the three HST LFs is a bold 
line.  Power-law fit and extension is solid line (see text). }
\end{figure}

\clearpage
\begin{figure}
\epsscale{0.5}
\plotone{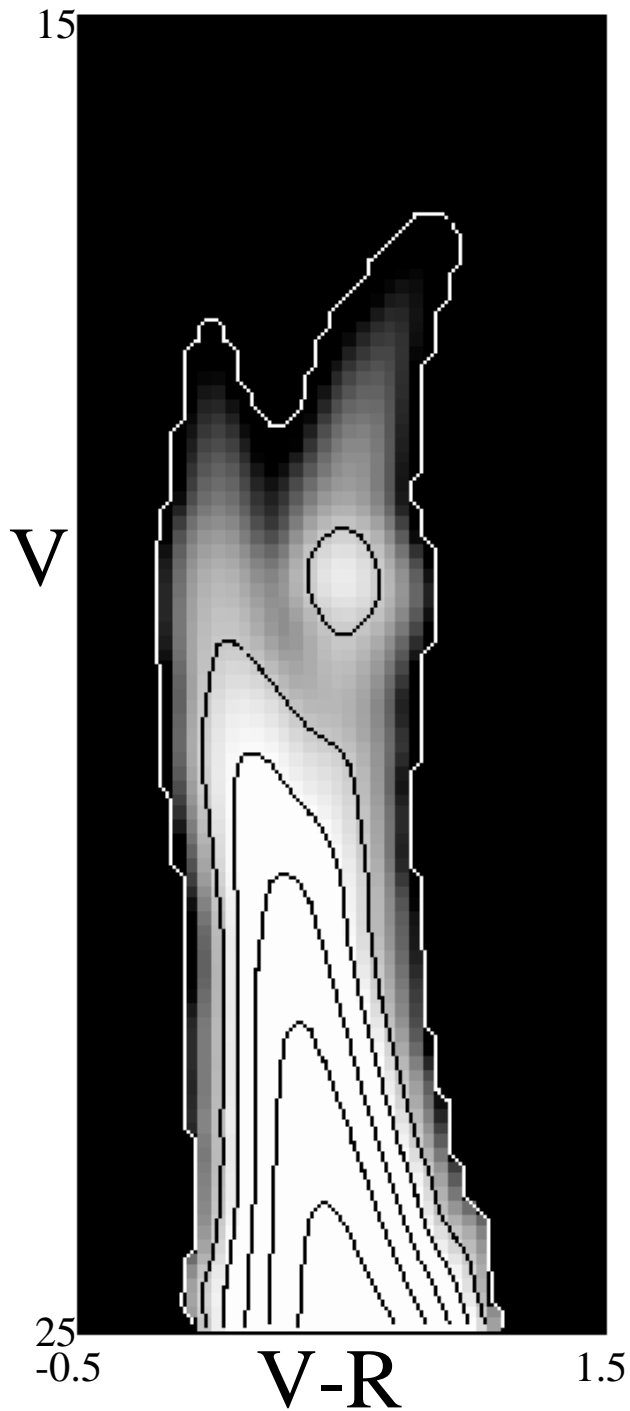}
\caption{The efficiency CMD; a Hess diagram (2-d histogram)
representing the number of stars
per 0.52 square degrees in the LMC as a function of $V$ and $(V-R)$.  
Intensity represents the number of stars.  Logarithmic contours
are overplotted in the diagram.  The lowest contour (white) represents 
1.0 dex stars (per $\Delta(V-R) = 0.05$, $\Delta V$ = 0.10 mag bin, and
0.52 square degree) while the other contours run from 3.5 to 5.5 dex
in steps of 0.5 dex.}
\epsscale{1.0}
\end{figure}

\clearpage
\begin{figure}
\plotone{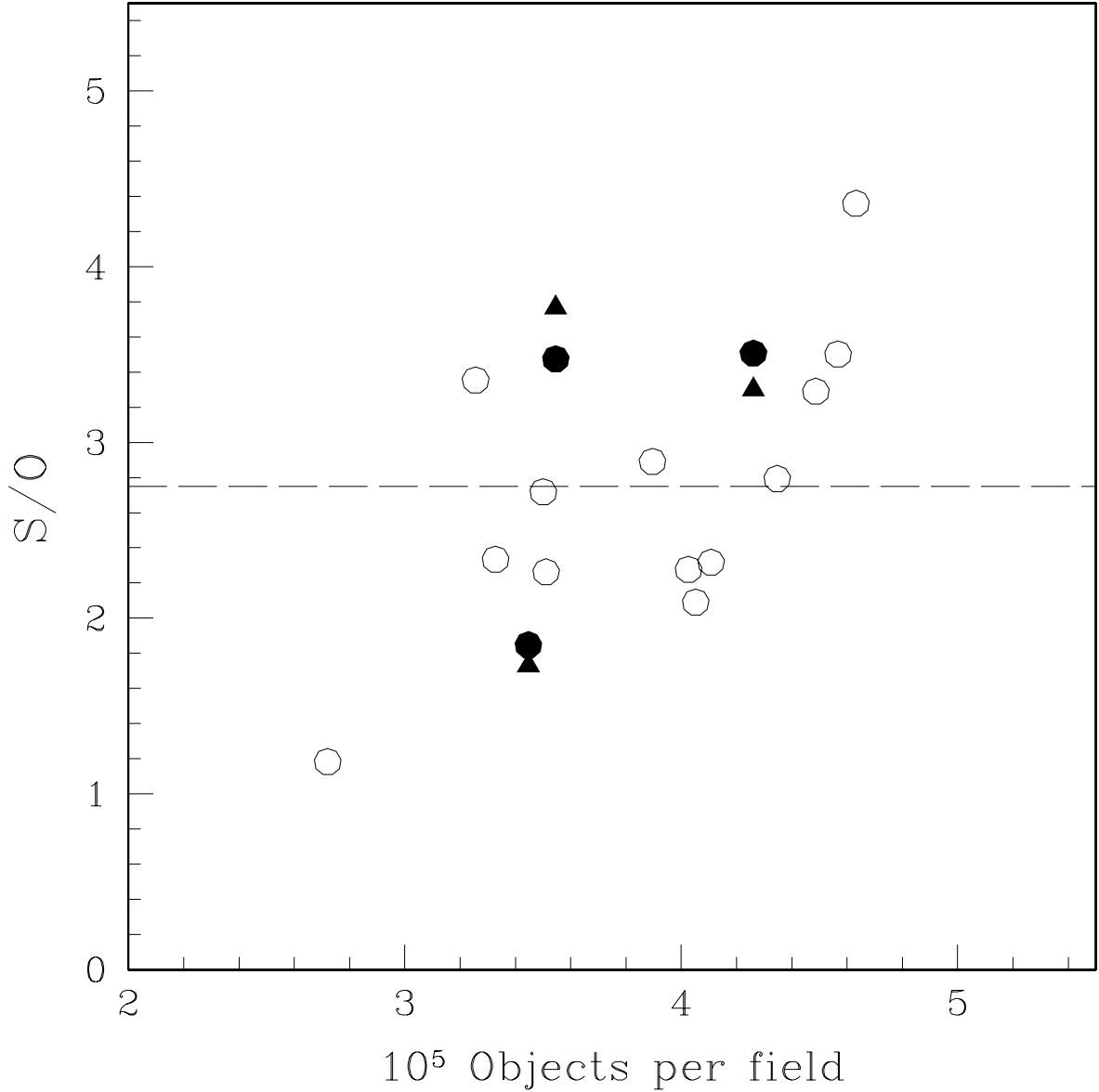}
\caption{Star/Object ratio (S/O) to the limiting magnitude
$V$ = 22 for 16 fields near the LMC bar as a function of the number
of MACHO objects per field.  Solid triangles are
calculated from HST/MACHO photometry comparisons for three fields (see text).
Solid circles show the fit values to a regression of these three S/O
values and surface brightness measurements.  The open circles
show the S/O ratios predicted by the surface brightness regression
for an additional 13 MACHO LMC fields.}
\end{figure}

\end{document}